\documentclass[fleqn,twoside]{article}
\usepackage{myespcrc2}

\usepackage{epsfig}
\usepackage{axodraw}

\usepackage{amssymb}

\newcommand{\bea}{\begin{eqnarray}}

\newcommand{\eea}{\end{eqnarray}}

\newcommand{\bean}{\begin{eqnarray*}}

\newcommand{\eean}{\end{eqnarray*}}

\newcommand{\nn}{\nonumber \\}

\newcommand{\be}{\begin{equation}}
\newcommand{\ee}{\end{equation}}

\def\spa#1.#2{\langle#1\,#2\rangle}
\def\spb#1.#2{[#1\,#2]}
\def\spab#1.#2.#3{\langle\mskip-1mu{#1} 
                  | #2 | {#3}]}

\def\spba#1.#2.#3{[\mskip-1mu{#1} 
                  | #2 | {#3}\rangle}

\def\spbb#1.#2.#3.#4{[\mskip-1mu{#1} 
                     | {#2} \ {#3} | {#4}]}

\def\spaa#1.#2.#3.#4{\langle\mskip-1mu{#1} 
                     | {#2} \ {#3} | {#4}\rangle}

\def\dea{\langle \ell \ d \ell \rangle}
\def\deb{[\ell \ d \ell]}
\def\dex{\int_0^1 dx}

\def\dedeb{[d \ell \ \partial_{\tilde{\ell}}]}

\def\nn{{\nonumber}}

\newbox\SlashedBox
\def\slashed#1{\setbox\SlashedBox=\hbox{#1}
\hbox to 0pt{\hbox to 1\wd\SlashedBox{\hfil/\hfil}\hss}#1}
\def\hboxtosizeof#1#2{\setbox\SlashedBox=\hbox{#1}
\hbox to 1\wd\SlashedBox{#2}}

\newbox\charbox
\newbox\slabox
\def\s#1{{      
        \setbox\charbox=\hbox{$#1$}
        \setbox\slabox=\hbox{$/$}
        \dimen\charbox=\ht\slabox
        \advance\dimen\charbox by -\dp\slabox
        \advance\dimen\charbox by -\ht\charbox
        \advance\dimen\charbox by \dp\charbox
        \divide\dimen\charbox by 2
        \raise-\dimen\charbox\hbox to \wd\charbox{\hss/\hss}
        \llap{$#1$}
}}

\newcommand\myTripleCut{
\begin{picture}(0,0)(0,0)
\SetScale{0.5}
\SetWidth{0.5}
\GOval(0,0)(30,30)(0){1}
\Line(-30,0)(-40,+10)
\Line(-30,0)(-40,-10)
\Line( 18,25)(18,40)
\Line( 18,25)(35,25)
\Line( 18,-25)(18,-40)
\Line( 18,-25)(35,-25)
\GOval(-30,0)(5,5)(0){0.7}
\GOval( 18, 25)(5,5)(0){0.7}
\GOval( 18,-25)(5,5)(0){0.7}
\DashLine(0,35)(0,-35){3}
\DashLine(0,0)(35,0){3}
\Text(-32, 0)[]{\tiny {${\cal A}_L(K)$}}
\Text( 25, 20)[]{\tiny {${\cal A}_M(K_2)$}}
\Text( 25, -20)[]{\tiny {${\cal A}_R(K_3)$}}
\Text(0,-25)[]{\tiny {$p$}}
\Text(0,+25)[]{\tiny {$p-K$}}
\Text(+30,0)[]{\tiny {$p+K_3$}}
\end{picture}
}

\begin{document}
\begin{titlepage}

\title{On Triple-Cut of Scattering Amplitudes}

\author{Pierpaolo Mastrolia \\ 
   {\it Institute of Theoretical Physics, University of Z\"urich, CH-8057}
}

\begin{abstract}
It is analysed the triple-cut of 
one-loop amplitudes in dimensional regularisation 
within spinor-helicity representation.
The triple-cut is defined as a difference of two double-cuts 
with the same particle contents, and
a same propagator carrying, respectively, causal 
and anti-causal prescription in each of the two cuts.
That turns out into an effective tool for extracting
the coefficients of three-point functions (and higher-point ones) 
from one-loop amplitudes. The phase-space integration is oversimplified
by using residues theorem to perform the integration
on the spinor variables, via the holomorphic anomaly, 
and a trivial integration on the Feynman parameter.
The results are valid for arbitrary values of dimensions.
\end{abstract}

\maketitle

\thispagestyle{empty}
\end{titlepage}

\section{Introduction}

It is a well known fact that any one-loop amplitude with massless
particles running in the loop, 
can be written, {\it via} standard 
Passarino-Veltman reduction, in terms of a basis of 
analytically known scalar integrals
\cite{tHooft:1979,Bern:1992em,Bern:1993kr,Tarasov:1996br},
called master integrals (MI). 
Such a basis consists of box-, triangle-, and bubble-diagrams
($I_4, I_3, I_2$ respectively), which
in four dimensions render the amplitude 
a combination of polylogarithms and rational terms. 

To compute any amplitude, it is therefore sufficient to compute
each of the rational coefficients entering that linear combination,
and the principle of unitarity-based methods, as proposed by
Bern, Dixon, Dunbar and Kosower \cite{Bern:1994zx},
is the exploitation of the unitarity-cuts\footnote{ 
In the following we equivalently use the terminology of {\it multiple} cut
and {\it multi-particle} cut.}
of each MI, for reading its
coefficient out of the amplitude.

Unitarity in four-dimension (4D) 
is sufficient to compute the polylogarithmic terms 
and the transcendental constants of one-loop amplitudes. 
By exploiting the analytic continuation of tree-amplitudes to complex 
spinors, initiated by Witten, Cachazo and Svrcek 
\cite{Witten:2003nn,Cachazo:2004kj},
and the properties of the complex integration 
\cite{Cachazo:2004zb,Cachazo:2004by,Cachazo:2004dr,Britto:2004nj},
new techniques have generalised the cutting rules.
On the one side, the {\it quadruple}-cut technique 
of Britto, Cachazo and Feng \cite{Britto:2004nc} 
yields the immediate computation of boxes' coefficient.
On the other side, the polylogarithmic structure related to 
box-, triangle- and 
bubble-functions can be detected by a {\it double}-cut 
and computed by a 
novel way of performing the phase-space integral
\cite{Britto:2005ha,Britto:2006sj},
introduced by Britto, Buchbinder, Cachazo and Feng,
in the context of Supersymmetry,
that with Britto and Feng, we further extended 
to deal with non-supersymmetric amplitudes,
which combines the extraction of residues in spinor 
variables and the integration over a Feynman parameter.

However, on general grounds, amplitudes in nonsupersymmetric theories, 
like QCD, suffer of rational ambiguities
that are not detected by the four-dimensional dispersive integrals.
Therefore, in the very recent past several groups have 
developed new techniques focusing on the separate 
computation of the rational term of one-loop amplitudes.
According to the combined {\it unitarity-bootstrap} approach, 
introduced by Bern, Dixon and Kosower in collaboration with Berger and Forde
\cite{UnitarityBootstrap},
the cut-containing terms computed by 4D-unitarity 
can provide corrective factors 
(due to factorisation constraints \cite{Bern:1995ix}) 
to a BCFW-like recurrence relation \cite{BCFW}
for the reconstruction of the rational part, from 
the rational part of lower-point amplitudes.
Xiao, Yang and Zhu have developed
an optimized tool by 
tailoring the Passarino-Veltman reduction on the integrals that are 
responsible of the rational part of scattering amplitudes
\cite{Xiao:2006vr}, 
giving rise to further refinements and new developments 
of algorithms for the tensor reduction of Feynman integrals 
like the {\it integrand decomposition} technique of Ossola, Papadopoulos 
and Pittau \cite{Ossola:2006us}, 
and the {\it form-factors} method of Binoth, Guillet and 
Heinrich \cite{Binoth:2006hk}. 

Alternatively, as it was realized by van Neerven \cite{vanNeerven:1985xr},
one can reconstruct the full amplitude  from
unitarity cuts in $D \ (=4-2\epsilon)$ dimensions. 

The unitarity-method introduced by Bern, Dixon, Dunbar,
Kosower and Morgan \cite{Bern:1994zx,Bern:1995db,SelfDual}
avoids the explicit evaluation of the phase-space integrals.
It rather relies on the channel-by-channel reconstruction of the 
loop-integrand, by lifting the $\delta^{(+)}$-functions to full propagators, 
after having exploited as much as possible the simplification 
due to the on-shell cut-conditions. 
Then, by means of conventional techniques for 
loop-integrals, one obtains the reduction of the of reconstructed-loop 
amplitude in terms of MI, enabling the extraction of their coefficients.

Brandhuber, McNamara, Spence and Travaglini \cite{Brandhuber:2005jw} 
combined that technique
of reconstructed-loop integrands with the generalised cutting rules, extending 
the effectiveness of the multiple-cuts, namely $quadruple$- and $triple$-cuts,
from four to $D$ dimensions.
In particular the $quadruple$-cuts of reconstructed amplitudes
yield the extraction of the coefficients of $n$-point MI, with $n \ge 4$; 
whereas the $triple$-cuts of reconstructed amplitudes
yield the extraction of the coefficients of $n$-point MI, with $n \ge 3$, 
(which do have three denominators to be cut), 
and also of those 2-point MI that (do not have them, but) 
come from the tensor reduction of triangle-functions.

Recently, together with Anastasiou, Britto, 
Feng and Kunszt \cite{ABFKM1,ABFKM2},
we have been able to extend, as well from the four dimensional case, 
the effectiveness of the true integration of the phase-space 
\cite{Britto:2005ha,Britto:2006sj},
for the computation of the two-particle cut in $D$ dimensions, 
by combining the extraction of residues in spinor variables and 
the parametric Feynman integration, convoluted 
with a further trivial parametric integration.
Double-cuts in $D$ dimensions can detect the coefficients of any $n$-point
MI, with $n \ge 2$, which are what needed for the computation of any scattering
amplitude (in absence of 0-point functions, the tadpoles). \\

In this letter, we present a new way for computing {\it triple-cuts} 
of dimensional regularised one-loop amplitudes.
It enables the direct extraction of
triangle- and higher-point-function coefficients from
any one-loop amplitude in arbitrary dimensions.
It combines the benefits of the double-cut integration of  
\cite{Britto:2005ha,Britto:2006sj,ABFKM1,ABFKM2}
and of the exploitation of the on-shell cut-conditions
\cite{Bern:1994zx,Bern:1995db,SelfDual,Brandhuber:2005jw},
through the idea of the inverse Cutkosky rule, already employed
by Anastasiou and Melnikov \cite{Cutkosky:1960sp,Anastasiou:2002yz}, 
to replace the third
on-shell $\delta$-function by the difference of two propagators,
\bea
(2 \pi i) \delta(p^2 - \mu^2) &\to& 
  {1 \over p^2 - \mu^2 + i0} 
- {1 \over p^2 - \mu^2 - i0} \ .
\nn 
\eea
 
That yields an effective disentangling of 
the algebraic reduction of the integrand, 
achieved by trivial spinor algebra (Schouten identities),
from the actual integrations which turn out to be oversimplified, 
when not trivialised.

Accordingly,
the triple-cut is treated as a difference of two double-cuts
with the same particle contents, and
a same propagator carrying respectively causal 
and anti-causal prescription in each of the two cuts.

The triple-cut phase-space for massless particle in $D$ dimensions
is written as a convolution of 
a four-dimensional triple-cut of massive particle, and an
integration over the corresponding mass parameter, which 
plays the role of a $(-2\epsilon)$-dimensional scale 
\cite{Mahlon:LoopMeasure}.

As for the double-cut \cite{ABFKM1,ABFKM2},
to perform the four-dimensional integration, we combine 
the method of spinor integration via the holomorphic anomaly
of massive phase-space integrals, and 
an integration over the Feynman parameter.
But, in the case of the triple-cut,
after Feynman parametrisation, by combining back the two double-cuts,
the parametric integration is reduced to the extraction of residues
to the branch points in correspondence of the zeroes
of a {\it standard quadratic function} (SQF'n) in the Feynman parameter.
It is that SQF'n, or better, 
its roots that carry the analytic information 
characterizing each master-integral, 
therefore determining its own generalised cuts, hereafter called
{\it master-cuts},
as it could be seen also 
from the seminal analysis by 't Hooft and Veltman of the generic scalar 
one-loop integrals \cite{tHooft:1979}.

The final integration over the dimensional scale parameter
is mapped directly to triple-cut of 
master integrals, possibly with shifted dimensions
\cite{Bern:1992em,Bern:1993kr,Tarasov:1996br}.

The method hereby developed can be considered as one more
dowel in the jigsaw of reconstructing any amplitude from its 
multiple generalised cuts
along the lines of 
the Feynman Tree Theorem \cite{TreeTheorem,Brandhuber:2005kd} 
and the Veltman Largest Time Equation \cite{LTE,Remiddi:1981hn}.
In this spirit, one could now compute
$n$-point ($n \ge 4$) coefficients from quadruple cuts,
three-point coefficients from triple-cuts, 
and two-point coefficients from double-cuts, by avoiding 
the conventional tensor reduction.
As it turns out,
given the decomposition of any amplitude in terms of MI, 
the coefficient of any $n$-point MI
can be recovered from the $n$-particle cut.
Obviously, that $n$-particle cut may detect as well higher-point MI, 
which will appear with different analytic structures, 
for they come from the zeroes of a SQF'n specific
of each diagram.

The triple-cut method hereby outlined 
can be applied to scattering amplitudes in gauge
theories, and in Gravity as well. In particular, in the latter case, 
it suitable for the analytic 
investigation of the so called ``no-triangle hypothesis'' 
for one-loop amplitudes in ${\cal N}=8$ four-dimensional Supergravity,
conjectured by Bern, Bjerrum-Bhor, and Dunbar, and already confirmed 
togeher with Ita, Perkins and Risager at the 6- and 7-point level 
\cite{NoTriangle}.

On the more speculative side,
we think that the characterization of master integrals in terms of (the branch
points corresponding to) the zeroes of the 
a standard function of the Feynman parameter
might lead to a deeper understanding of the decomposition of one-loop 
amplitudes in terms of basic scalar integrals; and,
together with the multi-particle cuts defined as iteration of (difference of)
one-particle cuts, possibly, of their recursive behaviour.

Before proceeding with the analysis of the triple-cut, let us recap the 
double-cut technique introduced in \cite{ABFKM1,ABFKM2}.

\section{Double-Cut}

We consider dimensional regularised one-loop amplitudes
with massless propagators
in the four-dimensional helicity  (FDH)
scheme, with external momenta living in four dimensions and the loop 
momentum living in a space with number of dimensions equal to
$D \ (=4-2\epsilon)$.  


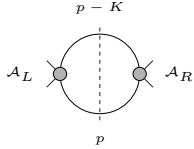
\begin{figure}[h]
$$
\begin{picture}(0,0)(0,0)
\SetScale{0.5}
\SetWidth{0.5}
\GOval(0,0)(30,30)(0){1}
\Line(-30,0)(-40,+10)
\Line(-30,0)(-40,-10)
\Line( 30,0)(40,+10)
\Line( 30,0)(40,-10)
\GOval(-30,0)(5,5)(0){0.7}
\GOval( 30,0)(5,5)(0){0.7}
\DashLine(0,35)(0,-35){3}
\Text(-30, 0)[]{\tiny {${\cal A}_L$}}
\Text( 30, 0)[]{\tiny {${\cal A}_R$}}
\Text(0,-25)[]{\tiny {$p$}}
\Text(0,+25)[]{\tiny {$p-K$}}
\end{picture}
$$
\caption{
Double-Cut.
}
\label{Fig:DoubleCut}
\end{figure}

The discontinuity \cite{Cutkosky:1960sp} 
of a generic one-loop amplitude 
in $D$-dimension is defined {\it via} the double-cut 
in Fig.\ref{Fig:DoubleCut}, corresponding to, 
\bea
{\cal M} &=& \int d^{4-2\epsilon}\Phi \ 
{\cal A}^{\rm tree}_L \times {\cal A}^{\rm tree}_R \ ,
\label{eq:doublecut}
\\
d^{4-2\epsilon}\Phi &=&
d^{4-2\epsilon} p \ 
\delta^{(+)}\left(p^2\right) 
\delta^{(+)}\left((K-p)^2\right)
\label{eq:deDp}
\eea
where 
$d^{4-2\epsilon}\Phi$ is the Lorentz invariant two-body phase-space 
in $D$ dimensions ($D$-LIPS),
${\cal A}^{\rm tree}_{L,R}$ are tree-level amplitudes, 
and $K$ is the total momentum
across the cut. 

Since the external momenta are
in four dimensions, we can decompose the loop momentum as 
$p = L +\vec{\mu}, $ where $L$ is a $4$-dimensional vector,
and $\vec{\mu}$ is its $(-2\epsilon)$-dimensional 
complement \cite{Mahlon:LoopMeasure,Bern:1995db}. 
As a consequence, the $D$-LIPS can be expressed as an integral over the 
dimensional scale $\mu^2$ of a four-dimensional 
$\mu^2$-dependent discontinuity, equivalently written as 
\bea
\int d^{4-2\epsilon} \Phi &=& 
\chi_K(\epsilon) \int_0^1 du \ u^{-1 - \epsilon} 
\int d^4 \Phi \ ,
\\
\chi_K(\epsilon) &=&
{(4\pi)^\epsilon \over \Gamma(-\epsilon)}
\left( {K^2 \over 4} \right)^{-\epsilon} \ ,
\\
u &=& {4 \mu^2 \over K^2} \ ,
\label{eq:udef}
\eea
\vspace*{-0.5cm}
\bea
\int d^4 \Phi \!=\! 
\int \! d^4 L \ 
\delta^{(+)}\!(L^2 - \mu^2) \ 
\delta^{(+)}\!((L - K)^2 - \mu^2) \ .
\ && 
\eea

In so doing, one can write the 
$D$-dimensional massless double-cut ${\cal M}$, essentially
as an integral in $u$ 
of a four dimensional (4D) massive double-cut, $\Delta$,
\bea
{\cal M} = 
\chi_K(\epsilon) \int_0^1 du \ u^{-1 - \epsilon} 
\ \Delta \ ,
\label{eq:DDCdef}
\eea
with the 4D-discontinuity defined as,
\bea
\Delta = \int d^4 \Phi \ 
{\cal A}^{\rm tree}_L(p) \ {\cal A}^{\rm tree}_R(p) \ .
\label{eq:Deltadef}
\eea

To perform the cut integration along the line of
\cite{Cachazo:2004dr,Britto:2004nj,Britto:2004nc,Britto:2005ha,Britto:2006sj},
the massive loop momentum $L$ is decomposed 
into a linear combination of a
light-like vector, $\ell_0$, 
and the time-like momentum-cut, $K$, \cite{ABFKM1,ABFKM2}
\bea
L = \ell_0 + z K \ , \qquad {\rm with\ \ } \ell_0^2=0 \ .
\label{eq:shift}
\eea
After the rescaling \cite{Cachazo:2004kj},
\bea
\ell_0^{a\dot{a}} = t \ \ell^{a} \tilde{\ell}^{\dot{a}} \ , 
\label{def:rescaling}
\eea
the 4D-massive double-cut integration  
appears to be parametrised as,
\bea
\int d^4 \Phi 
&=&
\int dz {\delta(z-z_0) \over (1-2z) K^2}
\int \dea \deb 
\times \nn \\ && \qquad
\int t^2 dt \ 
\delta \Bigg(t - {(1-2z) K^2 \over \spab \ell.K.\ell}\Bigg) 
\nn \\ 
&=&
\int dz \ \delta(z-z_0) 
\int {\dea \deb \over \spab \ell.K.\ell}
\times \nn \\ && \qquad
\int t \ dt \ 
\delta \Bigg(t - {(1-2z) K^2 \over \spab \ell.K.\ell}\Bigg) 
\label{eq:Phi4measure}
\\
z_0 &=& {1 - \sqrt{1-u} \over 2} \ ,
\label{eq:z0def}
\eea
where, 
$z_0$ is the proper root of the equation $z(1-z)K^2 - \mu^2=0$, as allowed by
the $\delta^{(+)}$-conditions.
One can see the similarities between the massive and massless phase-space in
four dimensions by comparing (\ref{eq:Phi4measure}) and 
(\ref{phi4massless}).

It is very important to notice that, due to the shift in (\ref{eq:shift}), 
the spinor integration becomes light-like as required by the method in 
\cite{Britto:2005ha,Britto:2006sj}.
Accordingly, by means of basic 
spinor algebra, namely by Schouten identities, one can 
disentangle the dependence over $|\ell\rangle$ and $|\ell]$, and
express the result of the $t$-integration as a combination 
of terms whose general form looks like,
\bea
\int t \ dt \ 
\delta \Bigg(t - {(1-2z)K^2 \over \spab \ell.K.\ell}\Bigg) 
{
{\cal A}^{\rm tree}_L(\ell,z,t) \ {\cal A}^{\rm tree}_R(\ell,z,t)
\over 
\spab \ell.K.\ell 
}
= && \nn \\
\sum_i \ 
{\cal G}_i\left(|\ell\rangle,z \right)
{ 
\spb \eta.\ell^{n}
\over 
\spab \ell.P_1.\ell^{n+1}
\spab \ell.P_2.\ell
} \ ,
\qquad &&
\label{eq:4Dgendeco}
\eea
where $P_i$ can either be equal to $K$, 
or be a linear combination of external vectors, 
which depends on $z$, coming from (off-shell) propagators;
and where ${\cal G}$'s depend solely on one spinor flavour, say 
$|\ell\rangle$ 
(and not on $|\ell]$), and may contain poles in $|\ell\rangle$ through 
factors like $1/\spa \ell.\Omega$ 
(with $|\Omega\rangle$ being a massless spinor, either associated to 
any of the external legs, say $|k_i\rangle$, 
or to the action of a vector on it, like $\s{P} |k_i\rangle$).

The explicit form of the vectors $P_1$ and $P_2$ 
in Eq.(\ref{eq:4Dgendeco}) is determining 
the nature of the 4D-double-cut, logarithmic or not,
and correspondingly the topology of the 
diagram which is associated to.
For easy of notation let us define the generic term in the {\it r.h.s.}
of Eq.(\ref{eq:4Dgendeco}), 
\bea
{\cal I}_i &=& 
{\cal G}_i\left(|\ell\rangle, z \right)
{ 
\spb \eta.\ell^{n}
\over 
\spab \ell.P_1.\ell^{n+1}
\spab \ell.P_2.\ell
}
\eea
Accordingly the 4D-discontinuity in Eq.(\ref{eq:Deltadef}) reads,
\bea
\Delta = 
\sum_i 
\int dz \ \delta(z-z_0) 
\int \dea \deb \ {\cal I}_i
\ .
\label{eq:DeltaForLater}
\eea

Let us distinguish among the two possibilities one encounters,
in carrying on the spinor integration of ${\cal I}_i$:
\begin{enumerate}
\item $P_1 = P_2 =K$ (momentum across the cut).
In this case, the result contains only  
the cut of a linear combination
bubble-functions with external momentum $K$, and 
dimensions which might be or not shifted from the original value, $D$.

\item $P_1 = K$, $P_2 \ne K$, or $P_1 \ne P_2 \ne K$. 
In this case, the result can contain 
the cut of a linear combination of 
$n$-point functions with $n\ge3$
and dimensions which might be, or not, shifted from 
the original value, $D$.
\end{enumerate}

Since, in this letter, 
we are mainly interested in triangle-functions (and higher-point ones), 
or better, in their coefficients, we will 
focus on case 2.

\subsection{Logarithmic Terms of 4D-Double-Cut}

Let us assume that 
either $P_1 = K$, $P_2 \ne K$ or $P_1 \ne P_2 \ne K$. 
In such a situation, one proceeds by introducing
a Feynman parameter, to write ${\cal I}_i$ as,
\bea
{\cal I}_i &=& 
(n+1) \dex \ 
(1-x)^n \ 
{\cal G}_i\left( |\ell\rangle, z \right)
{ 
\spb \eta.\ell^{n}
\over 
\spab \ell.R.\ell^{n+2} 
} 
\ , \quad
\eea
with 
\bea
\s{R} = x \s{P}_1 + (1-x) \s{P}_2 
\label{eq:Rdef}
\eea
We can then proceed with the spinor integration of ${\cal I}_i$
(the order of the integrations over the spinor variables and over 
the Feynman parameter can be exchanged).

First, one performs the integration over the $|\ell]$-variable
by parts, using \cite{Britto:2005ha}
\be
\deb {\spb \eta.\ell^n \over \spab \ell.P.\ell^{n+2}} 
=
{\dedeb \over (n+1)}
{
\spb \eta.\ell^{n+1} 
\over \spab \ell.P.\ell^{n+1} \spab \ell.P.\eta} \ .
\label{ibp}
\ee

Afterwards, the integration over the $|\ell\rangle$-variable, by 
using Cauchy's residues theorem in the fashion of the holomorphic anomaly
\cite{Cachazo:2004by,Cachazo:2004dr,Britto:2004nj}, 
%
%
yielding to,
\bea
&& {\cal F}_i = \int \dea \deb \ {\cal I}_i = 
\nn \\
&& = 
\dex \ (1-x)^n
\int \dea \dedeb
{
{\cal G}_i(|\ell\rangle, z) \ 
\spb \eta.\ell^{n+1} 
\over 
\spab \ell.R.\ell^{n+1} 
\spab \ell.R.\eta
} 
\nn \\
&&
= 
\dex \ (1-x)^n \ 
\bigg\{
{
{\cal G}_i(\s{R}|\eta], z)
\over 
(R^2)^{n+1}
}
\nn \\
&&
\qquad \quad +
\sum_j \lim_{\ell \to \ell_{ij}} 
\spa \ell.{\ell_{ij}}
{
{\cal G}_i(|\ell\rangle, z)) \ 
\spb \eta.\ell^{n+1} 
\over 
\spab \ell.R.\ell^{n+1} 
\spab \ell.R.\eta
}
\bigg\} 
\ , 
\label{ImplicitDoubleCut}
\eea
where $|\ell_{ij} \rangle$ are the simple poles of ${\cal G}_i$.

We may think to ${\cal F}_i$ in (\ref{ImplicitDoubleCut})
as decomposed into two pieces,
\bea
{\cal F}_i = {\cal F}_i^{{(1)}} + {\cal F}_i^{(2)}
\label{eq:FiDeco}
\eea
with
\bea
{\cal F}_i^{(1)} &\equiv&
\dex \ (1-x)^n \ 
{\cal G}_i(\s{R}|\eta], z) \ 
{1
\over 
(R^2)^{n+1}
} 
\ ,
\label{eq:ImplicitFi1def}
\eea
\bea
&& 
{\cal F}_i^{(2)} \equiv 
\dex \ (1-x)^n \ 
\nn \\
&&
\qquad \qquad \times
\sum_j \lim_{\ell \to \ell_{ij}} 
\spa \ell.{\ell_{ij}}
{
{\cal G}_i(|\ell\rangle, z)) \ 
\spb \eta.\ell^{n+1} 
\over 
\spab \ell.R.\ell^{n+1} 
\spab \ell.R.\eta
} \ ,
\label{eq:ImplicitFi2def}
\qquad
\eea

The expressions (\ref{eq:FiDeco}, 
\ref{eq:ImplicitFi1def}, \ref{eq:ImplicitFi2def}) are 
the key point for the triple-cut
construction, later discussed, therefore we will spend some words on it.

Let us observe that 
since $\s{R}$ is linear in $x$, as from Eq.(\ref{eq:Rdef}), 
$R^2$ is a quadratic function, the SQF'n, and can be written as,
\bea
R^2 = f(s_{ij},z) \ (x-x_1) \ (x-x_2) \ ,
\label{eq:R2def}
\eea
where $f$ may depend on $z$ and the invariants $s_{ij}=(k_i+k_j)^2$; and 
$x_{1,2}$ are the solutions of the equation $R^2=0$.

The key point is that $R^2$ 
is the signature of the master-cuts \cite{tHooft:1979}. 
More properly, its roots $x_1$ and $x_2$ are 
irrational functions of the kinematic scales, $s_{ij}$ and $\mu^2$,
specific for each diagram, and
allow to distinguish unequivocally among them.
In fact, 
the cuts of any scalar master-integral
are known from explicit calculation, and 
for triangle- and box-function one can see 
that the corresponding 4D-double-cut is proportional to the $\ln(x_1/x_2)$.

In particular the most general expression for $R^2$, accounting for both
$3m$-triangle and $4m$-box is a quadratic polynomial in $x$,
\bea
R^2 &=& a x^2 + 2 b x + c \ ,
\eea
with coefficients
\bea
 a &=& (P_1^2 - 2 P_1 \cdot P_2 + P_2^2) \ , \\ 
 b &=& P_1 \cdot P_2 - P_2^2 \ , \\
 c &=& P_2^2 \ ,
\eea
and zeroes at the values
\bea
x_{1,2} = {-b \pm \sqrt{b^2 - ac} \over a} \ . 
\eea

Given three vectors, $K_1, K_2, K_3$, bounded by momentum conservation,
$K_1 + K_2 + K_3 = 0$,
we can define the K\"allen $\lambda$-function
\bea
\lambda_{K_1,K_2,K_3} &=& 
 (K_1^2)^2 + (K_2^2)^2 + (K_3^2)^2 + 
 \nn \\ && 
- 2 K_1^2 K_2^2
- 2 K_1^2 K_3^2
- 2 K_2^2 K_3^2
\ . \quad  
\eea

We can, thus, write down the roots $x_{1,2}$ characterizing
triangle- and box-function, by using the following expressions,
according to the case. \\

\noindent
{\bf{$\bullet \ 3m$-Triangle}} 

\noindent
For a generic 3-point function with external legs labeled with $K_1, K_2, K_3$, 
and internal mass $\mu$, one has:
\bea
\s{P}_2 &=& \s{K}_1 \ ;
\\
\s{P}_1 &=& 
{K_3^2 + 2 z K_1 \cdot K_3 \over K_1^2} \s{K}_1 
+ (1-2z) \s{K}_3 
\ ; \quad 
\\
P_2^2 &=& K_1^2 \ ;
\\
P_1^2 &=& 
{1 \over K_1^2} 
\bigg(   {\mu^2 \over K_1^2} \ \lambda_{K_1,K_2,K_3}
       + K_2^2 K_3^2
\bigg) \ ;
\eea
\vspace*{-0.5cm}
\bea
2 P_1 \cdot P_2 &=& K_3^2 + K_2^2 - K_1^2 \ .
\eea

\noindent
{\bf{$\bullet \ 4m$-Box}} 

\noindent
For a generic 4-point function with external legs labeled 
with $K_1, K_2, K_3, K_4$ and internal mass $\mu$, one has:
\bea
\s{P}_1 &=& 
{K_1^2 - 2 z K_{12} \cdot K_1 \over K_{12}^2} \s{K}_{12} 
+ (1-2z) \s{K}_1 
\ ; \qquad
\\
\s{P}_2 &=& 
{K_4^2 + 2 z K_{12} \cdot K_4 \over K_{12}^2} \s{K}_{12} 
+ (1-2z) \s{K}_4 
\ ; \quad 
\\
P_1^2 &=& 
{1 \over K_{12}^2} 
\bigg(   {\mu^2 \over K_{12}^2} \ \lambda_{K_{12},-K_2,-K_1}
       + K_{12}^2 K_1^2
\bigg) \ ;
\\
P_2^2 &=& 
{1 \over K_{12}^2} 
\bigg(   {\mu^2 \over K_{12}^2} \ \lambda_{K_{12},K_3,K_4}
       + K_{12}^2 K_4^2
\bigg) \ ;
\eea
\vspace*{-0.5cm}
\bea
2 P_1 \cdot P_2 &=& 
{1 \over K_{12}^2} 
\Big[
  K_1^2 K_3^2 
+ K_2^2 K_4^2 
+ K_{12}^2 K_{41}^2
  \nn \\ && \qquad \ \ 
- 4 \mu^2 (K_1^2 + K_4^2 - K_{41}^2 )
\Big] \ .
\eea

Sub-cases for triangles and boxes 
with massless legs can be obtained by setting in the above expressions
the corresponding momentum-square to zero.

\vspace*{0.5cm}
The completion of the 4D-integration in (\ref{eq:DeltaForLater}), 
which reads,
\bea
\Delta = 
\sum_i 
\int dz \ \delta(z-z_0) 
\ {\cal F}_i
\ ,
\eea
can be achieved by merely substituting in the result 
of (\ref{eq:FiDeco})
the value $z=z_0$, given in (\ref{eq:z0def})
 - hereafter is understood the equivalence of
$z$ and $z_0$, and whenever $z$ appears, $z_0$ should be intended.

Finally, in order to get the the discontinuity in $D$ dimension 
(\ref{eq:DDCdef}), one should perform the very last integration over the 
dimensional parameter $u \ (= 4 \mu^2/K^2)$.
Indeed, the $u$-integral is not to be carried out 
explicitly: it can either be expressed 
in terms of shifted dimension master-cut with coefficients not depending
on $\epsilon$
\cite{Bern:1992em,Bern:1993kr,Bern:1995db,Tarasov:1996br,Brandhuber:2005jw};
or equivalently, as explained in \cite{ABFKM1,ABFKM2}, it
can be reduced {\it via}
recurrence relations (obtained by integration-by-parts identities), 
to master cut in $D$ dimensions and $\epsilon$-dependent coefficients.

As a last remark, we observe that
the 4D-massive discontinuity, which can be considered 
the kernel of the $D$-dimension integration (\ref{eq:DDCdef}), carries
all the main information about the decomposition in terms of master-cuts
\cite{tHooft:1979}.  
Due to the role played by the integration over the dimensional variable $u$,
the decomposition of the $D$-regularised cut-amplitude 
in terms of master-cuts in $D$ dimensions, 
stems from the decomposition of the 4D-massive cut-amplitude, in terms of
4D-massive master-cut.

\section{Triple-Cut}

\begin{figure*}[t]
\vspace*{0.5cm}
$$
\myTripleCut 
\hspace*{1.5cm} 
= \quad  
 {1 \over (2 \pi i)} \ 
\Bigg\{  
    \hspace*{1.0cm} 
\begin{picture}(0,0)(0,0)
\SetScale{0.5}
\SetWidth{0.5}
\GOval(0,0)(30,30)(0){1}
\Line(-30,0)(-40,+10)
\Line(-30,0)(-40,-10)
\Line( 18,25)(18,40)
\Line( 18,25)(35,25)
\Line( 18,-25)(18,-40)
\Line( 18,-25)(35,-25)
\GOval(-30,0)(5,5)(0){0.7}
\GOval( 18, 25)(5,5)(0){0.7}
\GOval( 18,-25)(5,5)(0){0.7}
\DashLine(0,35)(0,-35){3}
\Text(22, -5)[]{\tiny {$+i0$}}
%
\end{picture}
 \hspace*{1.0cm} - \hspace*{1.0cm} 
\begin{picture}(0,0)(0,0)
\SetScale{0.5}
\SetWidth{0.5}
\GOval(0,0)(30,30)(0){1}
\Line(-30,0)(-40,+10)
\Line(-30,0)(-40,-10)
\Line( 18,25)(18,40)
\Line( 18,25)(35,25)
\Line( 18,-25)(18,-40)
\Line( 18,-25)(35,-25)
\GOval(-30,0)(5,5)(0){0.7}
\GOval( 18, 25)(5,5)(0){0.7}
\GOval( 18,-25)(5,5)(0){0.7}
\DashLine(0,35)(0,-35){3}
\Text(22, -5)[]{\tiny {$-i0$}}
%
\end{picture}
\hspace*{1.0cm} 
\Bigg\}
$$
\caption{
Triple-cut in terms of two double-cuts, respectively with a causal propagator 
and an anti-causal propagator: ${\cal A}_L, {\cal A}_M$, and  
${\cal A}_R$ are tree level amplitudes,
respectively depending on the external momenta $K, K_2, K_3$.
}
\label{Fig:TripleCut}
\end{figure*}

The triple-cut of a generic one-loop amplitude 
in $D$-dimension is defined as 
\bea
{\cal N} &=& \int d^{4-2\epsilon} \Phi \ 
  \delta^{(+)}((p+K_3)^2) \ 
  {\cal A}_L
\ {\cal A}_M
\ {\cal A}_R \ ,
\label{eq:triplecut}
\eea
where: $\int d^{4-2\epsilon}\Phi$ 
is the two-body phase space defined for the double-cut in (\ref{eq:deDp});
$(p+K_3)$ is the momentum corresponding to the extra cut-propagator;
and ${\cal A}_{L,M,R}$ are the tree-level amplitudes forming the
one-loop pattern.

By using the inverse of Cutkowsky rule 
\cite{Cutkosky:1960sp,Anastasiou:2002yz}, to express
the $\delta^{(+)}((p+K_3)^2)$ as a difference of two scalar propagators with
opposite $i0$-prescription, one can write the triple-cut
\bea
{\cal N} &=& {1 \over (2 \pi i)}
 \int d^{4-2\epsilon} \Phi \ 
  {\cal A}_L
\ {\cal A}_M
\ {\cal A}_R 
\nn \\ 
&& \qquad \times
\left(
{1 \over (p+K_3)^2 + i0}
-
{1 \over (p+K_3)^2 - i0}
\right) 
\nn \\ 
&=&
{1 \over (2 \pi i)} ({\cal M}^{+} - {\cal M}^{-}) \ ,
\label{eq:tripleVSdubble}
\eea
as a difference of two double-cuts ${\cal M}^{\pm}$, 
with a same propagator carrying respectively a causal and anti-causal
$i0$-prescription in each of the two double-cuts, 
see Fig.\ref{Fig:TripleCut}
, where
\bea
{\cal M}^{\pm} 
&\equiv&
 \int d^{4-2\epsilon} \Phi \ 
  {\cal A}_L
  \times
{ 
\ {\cal A}_M
\ {\cal A}_R 
 \over (p+K_3)^2 \pm i0}
\nn \\
&=&
 \int d^{4-2\epsilon} \Phi \ 
  {\cal A}_L
  \times
{ 
\ {\cal A}_M
\ {\cal A}_R 
 \over (p+K_3)^2_\pm}
\ .
\eea
Hereafter, we keep track of the triple-cut propagator 
with the sub-index $\pm$.
In this form, one can deal with ${\cal M}^{+}$ and ${\cal M}^{-}$
as done in the previous section 
from Eq.(\ref{eq:DDCdef}) to Eq.(\ref{eq:ImplifitFi2def}),
but by taking care of 
the presence of the $\pm i0$.
Accordingly one has,
\bea
{\cal M}^{\pm} &=&  
\chi_K(\epsilon) \int_0^1 du \ u^{-1 - \epsilon} \ \Delta^{\pm} 
\label{eq:tiple:MvsDelta}
\\
&=&
\chi_K(\epsilon) \int_0^1 du \ u^{-1 - \epsilon} 
\sum_i 
\int dz \ {\delta(z-z_0)}
\ {\cal F}_i^{\pm}
\nn \\
&=&
\chi_K(\epsilon) \int_0^1 du \ u^{-1 - \epsilon} 
\int dz \ {\delta(z-z_0)}
\nn \\ && \qquad \times 
\sum_i 
\left(
{\cal F}_i^{(1,\pm)} + {\cal F}_i^{(2,\pm)}
\right) \ ,
\eea
with
\bea
{\cal F}_i^{(1,\pm)} &\equiv&
\dex \ (1-x)^n \ 
{\cal G}_i(\s{R}|\eta], z) \ 
{1
\over 
(R^2_{\pm})^{n+1}
} 
\ ,
\label{eq:tricut:F1pm}
\eea
\bea
&& 
{\cal F}_i^{(2,\pm)} \equiv 
\dex \ (1-x)^n \ 
\nn \\
&&
\qquad \qquad \times
\sum_j \lim_{\ell \to \ell_{ij}} 
\spa \ell.{\ell_{ij}}
{
{\cal G}_i(|\ell\rangle, z)) \ 
\spb \eta.\ell^{n+1} 
\over 
\spab \ell.R.\ell^{n+1}_\pm 
\spab \ell.R.\eta_\pm
} \ .
\qquad
\label{eq:tricut:F2pm}
\eea

In this fashion, the $D$-dimensional massless triple-cut, ${\cal N}$, 
as well as for the double-cut, can be interpreted as a 
$u$-integral of a 4D-massive triple-cut, $\Theta$,
\bea
{\cal N} &=&  
\chi_K(\epsilon) \int_0^1 du \ u^{-1 - \epsilon} \ \Theta \ ,
\label{eq:triple:DCUTvs4CUT}
\eea
where,
\bea
\Theta &=& {1 \over (2 \pi i)} ( \Delta^{+} - \Delta^{-} ) 
\nn \\ 
&=&
{1 \over (2 \pi i)}
\int dz \ {\delta(z-z_0)}
\nn \\ && \times
\sum_i 
\left(
  {\cal F}_i^{(1,+)} 
+ {\cal F}_i^{(2,+)} 
- {\cal F}_i^{(1,-)} 
- {\cal F}_i^{(2,-)} 
\right) .
\quad
\label{eq:thetadef}
\eea

To reach the form for ${\cal F}_i^{(1,\pm)}$ and ${\cal F}_i^{(2,\pm)}$ 
given in (\ref{eq:tricut:F1pm}) and (\ref{eq:tricut:F2pm}) respectively,
the Feynman parametrization should involve the extra cut-denominator
$(p + K_3)^2$, which, after the shift (\ref{eq:shift}) and the rescaling 
(\ref{def:rescaling}) has become,
\bea
(p + K_3)^2 \to {t \over (1-2z)} \spab \ell.Q.\ell \ , 
\eea
with 
\bea
\s{Q} = (1-2z) \s{K}_3 + {K_3^2 + 2z K \cdot K_3 \over K^2} \s{K}
\ .
\eea
In other words, the spinor algebra should be properly tailored 
to achieve a decomposition such that $\s{R}$ appears to be defined as a 
combination of two vectors, out of which one is $\s{Q}$,
\bea
\s{R} = x \s{P}_1 + (1-x) \s{Q} \ .
\label{eq:triple:Rdef}
\eea
In fact, $R$-type terms that after the Feynman parametrization do not 
contain $Q$, therefore without any memory of the $i0$-prescription, 
will just vanish from the triple-cut, once the
two double-cut-like contributions, ${\cal M}^{\pm}$, will be combined back. 

Before performing the $x$-integration, as one would do for the computation of 
a double-cut, in this case one can look at the full form of the triple-cut 
(\ref{eq:tripleVSdubble}) in terms
of ${\cal F}_i^{(1,\pm)}$ and ${\cal F}_i^{(2,\pm)}$,
\bea
{\cal N} &=& 
\chi_K(\epsilon) \int_0^1 du \ u^{-1 - \epsilon} 
\int dz \ {\delta(z-z_0)} \ \Theta 
\nn \\ 
&=&
\chi_K(\epsilon) \int_0^1 du \ u^{-1 - \epsilon} 
\int dz \ {\delta(z-z_0)}
\nn \\ && 
\times
\sum_i 
\Big\{
  \delta {\cal F}_i^{(1)} + \delta {\cal F}_i^{(2)}
\Big\}
\ ,
\label{eq:triplecutfinal}
\eea
with
\bea
\delta {\cal F}_i^{(1)} 
&\equiv&
  {1 \over (2 \pi i)} \left({\cal F}_i^{(1,+)} - {\cal F}_i^{(1,-)}\right)
\\
\delta {\cal F}_i^{(2)} &\equiv&
  {1 \over (2 \pi i)} \left({\cal F}_i^{(2,+)} - {\cal F}_i^{(2,-)}\right)
\ . 
\eea
The integration over the Feynman parameter, 
to be still performed both in
$\delta {\cal F}_i^{(1)}$ and in $\delta {\cal F}_i^{(2)}$, is 
frozen by the presence of a $\delta$-function, as follows,
\bea
\delta {\cal F}_i^{(1)} 
&=&
\dex \ (1-x)^n \ 
{\cal G}_i(\s{R}|\eta], z) \ 
\delta \Big((R^2)^{n+1}\Big)
\ , \qquad 
\label{deF1genericfinal}
\eea
\bea
\delta {\cal F}_i^{(2)} =
&&
\hspace*{-0.2cm}
\sum_j \lim_{\ell \to \ell_{ij}} 
\spa \ell.{\ell_{ij}} \ 
{\cal G}_i(|\ell\rangle, z)) \ \spb \eta.\ell^{n+1} 
   \nn \\ &&
\hspace*{-0.5cm}
\times
\dex \ (1-x)^n \ 
\delta \Big(\spab {\ell_{ij}}.R.{\ell_{ij}}^{n+1}
            \spab {\ell_{ij}}.R.\eta \Big) .
\qquad
\label{deF2genericfinal}
\eea
The expressions
(\ref{eq:triplecutfinal}, \ref{deF1genericfinal}, \ref{deF2genericfinal})
represent the final form of a generic three-particle cut.

\section{Examples}
\label{sec:Examples}

In the followings, we show some examples
of the application of the triple-cut method.
In Sec.\ref{sec:Scalar1mTriangle}
and Sec.\ref{sec:Scalar0mBox},
we compute the triple-cut of 
two master-integrals, namely 
the $1m$-triangle, and the $0m$-box; 
in Sec.\ref{sec:Linear1mTriangle} and Sec.\ref{sec:Linear0mBox}
we compute the triple-cut of a linear triangle and a linear box
integrals, respectively, 
to extract the coefficients of the $1m$-triangle, and the $0m$-box,
in agreement with the results in the literature.

\subsection{Scalar $1m$-Triangle}
\label{sec:Scalar1mTriangle}

\begin{figure}[h]
$$
\begin{picture}(0,0)(0,0)
\SetScale{0.5}
\SetWidth{0.5}
\Line(-20,0)(20,-20)
\Line(20,-20)(20,20)
\Line(20,20)(-20,0)
\Line(20,-20)(30,-30)
\Line(-20,0)(-30,10)
\Line(-20,0)(-30,-10)
\Line(20,20)(30,30)
\DashLine(0,30)(0,-30){3}
\DashLine(0,0)(28,0){3}
\Text(-20,-10)[]{\tiny {$1$}}
\Text(-20, 10)[]{\tiny {$2$}}
\Text( 20, 20)[]{\tiny {$3$}}
\Text( 20,-20)[]{\tiny {$4$}}
%
\Text( 0, 20)[]{\tiny {$L_2$}}
\Text( 20, 0)[]{\tiny {$L_3$}}
\Text( 0,-20)[]{\tiny {$L_4$}}
\end{picture}
$$
\caption{
Triple-cut of a $1m$-Triangle.
}
\label{fig:TripleCut1mTriangle}
\end{figure}
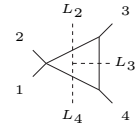

We consider the scalar integral represented in 
Fig.\ref{fig:TripleCut1mTriangle}, and associated to 
the triple-cut,
\bea
{\cal N}_{12|3|4} &=& 
\int d^D \Phi \  
\delta((L_2 - k_3)^2 - \mu^2)
\nn \\
&=&
\int d^D \Phi \  
\delta(2 L_2 \cdot k_3 )
\nn \\
&\equiv&
{1 \over (2 \pi i)} \int d^D \Phi \  
\Bigg\{
  {1 \over (2 L_2 \cdot k_3 ) + i 0}
  \nn \\ && \hspace*{2.5cm}
- {1 \over (2 L_2 \cdot k_3 ) - i 0}
\Bigg\}
\nn \\
&\equiv&
{1 \over (2 \pi i)} \int d^D \Phi \  
\Bigg\{
  {1 \over (2 L_2 \cdot k_3 )_{+}}
- {1 \over (2 L_2 \cdot k_3 )_{-}}
\Bigg\}
\nn \\
&\equiv& 
{1 \over (2 \pi i)} \Big\{ {\cal M}_{12}^{(+)} - {\cal M}_{12}^{(-)} \Big\} 
\nn \\
&=&
\chi_{12}(\epsilon) 
\int_0^1 du \ u^{-1-\epsilon} 
\ 
{1 \over (2 \pi i)} \Big\{ \Delta_{12}^{(+)} - \Delta_{12}^{(-)} \Big\} 
%
\nn \\ 
\label{scalar:triple:T12-3-4:def}
\eea
In so doing we have defined the triple-cut as a difference 
(modulo the overall factor ${1 \over (2 \pi i)}$)
of two double-cuts,
each carrying memory of its own $i0$-prescription.

The integration of the 4D-discontinuities, $\Delta^{\pm}$, can proceed 
as well as for a double-cut.
By defining, 
\bea
 \s{Q} &=& (1-2z) \s{k}_3 - z \s{K}_{12} \nn \\
   &=& (1-2z) \s{k}_3 + z \s{K}_{34} \nn \\
   &=& (1-z) \s{k}_3 + z \s{k}_4 
\ ,
\eea
after the shift (\ref{eq:shift}) and the rescaling (\ref{def:rescaling}) 
with cut-momentum, $K_{12} \ (= - K_{34})$, one has 
\bea
\Delta^{(\pm)}_{12} 
&=&
- (1-2z)
\int \dea \deb
{1 \over 
\spab \ell.K_{34}.\ell
\spab \ell.Q.\ell_\pm
} 
\nn \\
&=& 
- (1-2z)
\dex 
\int \dea \deb {1 \over \spab \ell.R.\ell^2_\pm} \ . 
\eea
with
\bea
\s{R} = x \s{Q} + (1-x) \s{K}_{34}
\eea
Then one proceeds with
\bea
\Delta^{(\pm)}_{12} 
&=& 
- (1-2z)
\dex 
\int \dea \dedeb 
{\spb \eta.\ell 
\over 
\spab \ell.R.\ell_\pm
\spab \ell.R.\eta
}
\nn \\ 
\eea
which has a pole at $|\ell\rangle = \s{R}|\eta]$.
By taking its residue, one gets
\bea
\Delta^{(\pm)}_{12} 
&=&
- (1-2z) \dex {1 \over R^2_\pm} \ ,
\label{scalar:Triple:Delta12def}
\eea
with
\bea
R^2 = z(1-z) \ s_{12} \ (x - x_1) (x - x_2) \ ,
\eea
and
\bea
x_1 &=& {1\over z} \ = \ \frac{2 \left(1 + \sqrt{1-u}\right)}{u} 
\nn \\
x_2 &=& {1 \over 1-z} \ = \  \frac{2 \left(1 - \sqrt{1-u}\right)}{u} \ . 
\label{Rroots4Triangle1m}
\eea
We remark that $x_{1,2}$ can be considered as the {\it characters} 
of the $1m$-triangle.

We use at this stage the definition of the $\delta-$function, 
yielding to the following expression for the 4D-massive triple-cut,
\bea
\Theta_{12|3|4} 
&=& {1 \over (2 \pi i)} \Big\{ \Delta_{12}^{(+)} - \Delta_{12}^{(-)} \Big\} 
\nn \\
&=& - (1-2z) \int dx \ \delta( R^2 ) 
\nn \\
&=& - {(1-2z) \over z (1-z) s_{12}} 
\int dx \ \delta \Big( (x-x_1) (x-x_2) \Big)
\nn \\
&=&
- {(1-2z) \over z (1-z) s_{12}} 
\Bigg\{
{ 1 \over |x_1 - x_2| }
+
{ 1 \over |x_2 - x_1 | }
\Bigg\}
\nn \\
&=&
- {2 (1-2z) \over z (1-z) \ s_{12} \ |x_1 - x_2|} 
\nn \\
&=& - {2 \over s_{12}} \ ,
\eea
which is independent of $u$, therefore of $\mu^2$.

Finally, the complete $D$-dimensional triple-cut of a $1m$-triangle $I_{3,1m}$ 
reads,
\bea
{\cal N}_{12|3|4} 
&=&
\chi_{12}(\epsilon) 
\int_0^1 du \ u^{-1-\epsilon} \ 
\Theta_{12|3|4} 
\nn \\ 
&=&
\chi_{12}(\epsilon) 
\int_0^1 du \ u^{-1-\epsilon} \ 
{ (- 2) \over s_{12}} \ .
\label{finalTripleCutTriangle1m}
\eea

\subsection{Scalar $0m$-Box}
\label{sec:Scalar0mBox}

\begin{figure}[h]
$$
\begin{picture}(0,0)(0,0)
\SetScale{0.5}
\SetWidth{0.5}
\Line(-20,-20)(20,-20)
\Line(20,-20)(20,20)
\Line(20,20)(-20,20)
\Line(-20,20)(-20,-20)
\Line(-20,-20)(-30,-30)
\Line(20,-20)(30,-30)
\Line(20,20)(30,30)
\Line(-20,20)(-30,30)
\DashLine(0,30)(0,-30){3}
\DashLine(0,0)(28,0){3}
\Text(-20,-20)[]{\tiny {$1$}}
\Text(-20, 20)[]{\tiny {$2$}}
\Text( 20, 20)[]{\tiny {$3$}}
\Text( 20,-20)[]{\tiny {$4$}}
\Text(-20, 0)[]{\tiny {$L_1$}}
\Text( 0, 20)[]{\tiny {$L_2$}}
\Text( 20, 0)[]{\tiny {$L_3$}}
\Text( 0,-20)[]{\tiny {$L_4$}}
\end{picture}
$$
\caption{Triple-cut of a $0m$-Box.}
\label{fig:TripleCut0mBox}
\end{figure}
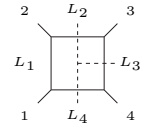

We consider the scalar integral represented in 
Fig.\ref{fig:TripleCut0mBox}, and associated to 
the triple-cut,
\bea
N_{12|3|4} &=& 
\int d^D \Phi \  
\delta((L_2 - k_3)^2 - \mu^2) {1 \over (L_2 + k_2) - \mu^2}
\nn \\
&=&
\int d^D \Phi \  
\delta(2 L_2 \cdot k_3 ) {1 \over (2 L_2 \cdot k_2)}
\nn \\
&=&
\chi_{12} \int_0^1 du \ u^{-1 - \epsilon} \ \Theta_{12|3|4} 
\nn \\
&=&
\chi_{12} \int_0^1 du \ u^{-1 - \epsilon} \ 
{1 \over (2 \pi i)} \Big\{ \Delta_{12}^{(+)} - \Delta_{12}^{(-)} \Big\} 
\eea
with 
\bea
\Delta_{12}^{(\pm)} &=&
\int d^4\Phi 
{1 \over (2 L_2 \cdot k_3 )_\pm \ (2 L_2 \cdot k_2)}
\ . 
\eea
Since one has,
\bea
2 L_2 \cdot k_3 &=& {t \over (1-2z)} \spab \ell.Q_1.\ell \ ,
\\
2 L_2 \cdot k_2 &=& {t \over (1-2z)} \spab \ell.Q_2.\ell \ ,
\eea
having defined 
\bea
\s{Q}_1 &\equiv&
(1-2z) \s{k}_3  + z \s{K}_{34} = 
(1-z) \s{k}_3  + z \s{k}_4 \ ,
\\
\s{Q}_2 &\equiv&  
(1-2z) \s{k}_2  + z \s{K}_{12} = 
(1-z) \s{k}_2  + z \s{k}_1 \ ,
\eea
after the shift (\ref{eq:shift}) and the rescaling (\ref{def:rescaling}) 
with cut-momentum, $K_{12} \ (= - K_{34})$,
one can write,
\bea
\Delta_{12}^{(\pm)}
&=&
\int dz {\delta(z-z_0) \over s_{12}}(1-2z)
\int  { \dea \deb \over \spab \ell.Q_1.\ell_\pm \ \spab \ell.Q_2.\ell }
\ . \nn \\
\eea
We give as understood the trivial $z$-integration, 
and perform the spinor integration
as for a double-cut,
\bea
\Delta_{12}^{(\pm)}
&=&
{(1-2z) \over s_{12}}
\int \dea \deb { 1 \over \spab \ell.Q_1.\ell_\pm \ \spab \ell.Q_2.\ell }
\nn \\
&=&
- 
{(1-2z) \over s_{12}}
\dex 
\int \dea \deb {1 \over \spab \ell.R.\ell^2_\pm} \ ,
\nn \\
&=&
- {(1-2z) \over s_{12}}
\dex {1 \over R_\pm^2} \ ,
\eea
with
\bea
\s{R} &=& x \s{Q}_2 + (1-x) \s{Q}_1 \ , 
\\
R^2 &=& 
  s_{23} (1 - 2 z)^2 x^2 
- s_{23} (1 - 2z)^2 x 
+ s_{12} z (1-z)
\nn \\ 
&=&
s_{23} (1 - 2 z)^2 \ (x-y_1) (x-y_2) \ ,
\eea
where
\bea
y_{1,2} = \frac{1}{2} \left(1 \pm \frac{\sqrt{A u+1}}{\sqrt{1-u}}\right)
\ , 
\qquad
A = { s_{13} \over s_{23} } \ . 
\label{Rroots4Box}
\eea
We can therefore write the 4D-massive triple-cut as,
\bea
\Theta_{12|3|4} &=& 
{1 \over (2 \pi i)} \Big\{ \Delta_{12}^{(+)} - \Delta_{12}^{(-)} \Big\} 
\nn \\
&=&
- {(1-2z) \over s_{12}}
\int dx \ \delta(R^2) 
\nn \\
&=&
- { 1 \over s_{12} s_{23} \ (1 - 2 z) }
\int dx \ \delta \Big( (x-y_1) (x-y_2) \Big) 
\nn \\
&=&
- { 2 \over s_{12} s_{23} \ (1 - 2 z) \  |y_1 - y_2|} 
\nn \\
&=&
- { 2 \over s_{12} s_{23} \ \sqrt{1+Au}} 
\ ,
\eea
which constitutes the integrand of the 
$D$-dimension triple-cut
of a $0m$-box function  $I_{4,0m}$, finally reading as,
\bea
{\cal N}_{12|3|4} 
&=&
\chi_{12}(\epsilon) 
\int_0^1 du \ u^{-1-\epsilon} \ 
\Theta_{12|3|4} 
\nn \\ 
&=&
\chi_{12}(\epsilon) 
\int_0^1 du \ u^{-1-\epsilon} \ 
{ (-2) \over s_{12} s_{23} \ \sqrt{1+Au}} 
\ . \quad  
\label{finalTripleCutBox0m}
\eea

\subsection{Linear Triangle}
\label{sec:Linear1mTriangle}

We consider the linear triangle integral appearing in 
Eqs.(3.45--3.47) of \cite{Brandhuber:2005jw},
\bea
C^\nu &=& \int d^4 L_2 \ d^{-2 \epsilon}\mu \ 
{\mu^2 L_2^\nu \over
D_1 \ D_2 \ D_3
}
\eea
where
\bea
D_1 &=&  L_2^2 - \mu^2 \ ; \\
D_2 &=& (L_2 + K_{12})^2 - \mu^2 \ ; \\
D_3 &=& (L_2 - k_3)^2 - \mu^2 \ . 
\eea

\begin{figure}[h]
\vspace*{0.5cm}
$$
\begin{picture}(0,0)(0,0)
\SetScale{0.5}
\SetWidth{0.5}
\GOval(0,0)(27,27)(0){1}
%
%
\Line(20,-20)(30,-30)
\Line(20,20)(30,30)
%
%
\Line(-27,0)(-37,10)
\Line(-27,0)(-37,-10)
\DashLine(0,32)(0,-32){3}
\DashLine(0,0)(30,0){3}
\Text(-23,-10)[]{\tiny {$1$}}
\Text(-23, 10)[]{\tiny {$2$}}
\Text( 20, 20)[]{\tiny {$3$}}
\Text( 20,-20)[]{\tiny {$4$}}
%
\Text( 0, 20)[]{\tiny {$L_2$}}
\Text( 20, 0)[]{\tiny {$L_3$}}
\Text( 0,-20)[]{\tiny {$L_4$}}
\end{picture}
\hspace*{1cm} = 
c_3 \hspace*{1cm} 
\begin{picture}(0,0)(0,0)
\SetScale{0.5}
\SetWidth{0.5}
\Line(-20,0)(20,-20)
\Line(20,-20)(20,20)
\Line(20,20)(-20,0)
\Line(20,-20)(30,-30)
\Line(-20,0)(-30,10)
\Line(-20,0)(-30,-10)
\Line(20,20)(30,30)
\DashLine(0,30)(0,-30){3}
\DashLine(0,0)(28,0){3}
\Text(-20,-10)[]{\tiny {$1$}}
\Text(-20, 10)[]{\tiny {$2$}}
\Text( 20, 20)[]{\tiny {$3$}}
\Text( 20,-20)[]{\tiny {$4$}}
%
\Text( 0, 20)[]{\tiny {$L_2$}}
\Text( 20, 0)[]{\tiny {$L_3$}}
\Text( 0,-20)[]{\tiny {$L_4$}}
\end{picture}
$$
\caption{Triple-cut of a linear triangle in terms of the master triple-cut.}
\label{fig:TripleCutLinearTriangle}
\end{figure}
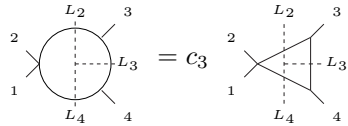

In particular let us consider the spinor sandwich 
$\spab 1.C.2$ which appears in Eq.(3.43) of \cite{Brandhuber:2005jw},
and whose result, obtained by PV-reduction, reads
\bea
\spab 1.C.2 = \spab 1.3.2 \ \lambda \ ,
\eea
with 
\bea
\lambda = J_3 - {2 \over s_{12}} J_2 \ ,
\eea
(overall factors understood, see \cite{Brandhuber:2005jw} for details).

Let us reconstruct the coefficient of $J_3$ from the triple-cut integration,
as depicted in Fig.\ref{fig:TripleCutLinearTriangle}.

We begin with,
\bea
{\cal N}_{12|3|4} &=& 
\int d^D \Phi \  
\delta((L_2 - k_3)^2 - \mu^2) \ 
\mu^2 \ 
\spab 1.L_2.2
\nn \\
&=&
\int d^D \Phi \  
\delta(2 L_2 \cdot k_3 ) \ 
\mu^2 \ 
\spab 1.L_2.2
\nn \\
&=&
\chi_{12} \int_0^1 du \ u^{-1 - \epsilon} \ 
\mu^2 \ \Theta_{12|3|4} 
\label{LinTri:DCUTvs4CUT}
\eea
where
\bea
\Theta_{12|3|4}  &\equiv&
{1 \over (2 \pi i)} \Big\{ \Delta_{12}^{(+)} - \Delta_{12}^{(-)} \Big\} 
\eea
with 
\bea
\Delta_{12}^{(\pm)} &=&
\int d^4\Phi 
{
\spab 1.L_2.2
 \over (2 L_2 \cdot k_3 )_\pm 
}
\nn \\
&=& 
\int dz \ \delta(z-z_0) \ s_{12}(1-2z)^2
 \nn \\ && \quad \times 
\int \dea \deb
{
\spab 1.\ell.2
 \over 
\spab \ell.K_{12}.\ell^2 \ 
\spab \ell.Q_1.\ell_\pm 
}
\eea
and 
\bea
 \s{Q}_1 &=& (1-2z) \s{k}_3 - z \s{K}_{12} \nn \\
   &=& (1-2z) \s{k}_3 + z \s{K}_{34} \nn \\
   &=& (1-z) \s{k}_3 + z \s{k}_4 
\ ,
\eea
as in the case of the scalar triangle.

We give as understood the $z$-integration and proceed with 
the spinor integration as for a double-cut.
\bea
\Delta_{12}^{(\pm)}
&=&
 s_{12}(1-2z)^2
\int \dea \deb
{
\spa 1.\ell \spb \ell.2
 \over 
\spab \ell.K_{12}.\ell^2 \ 
\spab \ell.Q_1.\ell_\pm 
} 
\nn \\
&=&
 s_{12}(1-2z)^2
\dex \ 
2 (1-x)
 \nn \\ && \times
\int \dea \deb
{
\spa 1.\ell \spb \ell.2
 \over 
\spab \ell.R.\ell_\pm^3
}
\eea
where
\bea 
\s{R} = x \s{Q}_1 + (1-x) \s{K}_{34} \ . 
\eea
We proceed with,
\bea
\Delta_{12}^{(\pm)}
&=&
 s_{12}(1-2z)^2
\dex \ 
(1-x)
 \nn \\ && \times
\int \dea \dedeb
{
\spa 1.\ell \spb \ell.2^2
 \over 
\spab \ell.R.\ell_\pm^2 
\spab \ell.R.2
}
\eea
which has a pole at $|\ell\rangle = \s{R}|2]$, therefore
\bea
\Delta_{12}^{(\pm)}
&=&
 s_{12}(1-2z)^2
\dex \ 
{(1-x) \spab 1.R.2 \over (R^2_\pm)^2}
\ , 
\eea
where
\bea
R^2 &=& z(1-z) \ s_{12}  (x - x_1) (x - x_2) \ .
\eea
and $x_{1,2}$ given in (\ref{Rroots4Triangle1m}).
$\Delta_{12}^{(\pm)}$ can be written as,
\bea
\Delta_{12}^{(\pm)}
&=&
 s_{12} \ \spab 1.3.2 \ (1-2z)^3 
\dex \ 
{(1-x) x \over (R^2_\pm)^2} \ ,
\nn \\
&=&
{ \spab 1.3.2 \ (1-2z)^3 
\over s_{12} \  z^2(1-z)^2
}
\dex \ 
{(1-x) x \over (D^2_\pm)^2} \ ,
\eea
where
\bea
D^2 = (x - x_1) (x - x_2) \ .
\eea
We can use the following identity,
\bea
{d \over dx} {1 \over D^2} 
&=& - { 2x - (x_1 + x_2) \over (D^2)^2}
\eea
to write
\bea
\Delta_{12}^{\pm}
&=&
- { \spab 1.3.2 \ (1-2z)^3 
\over s_{12} \  z^2(1-z)^2
}
 \nn \\ && \quad \times
\dex \ 
{(1-x) x \over 2x - (x_1 + x_2)} \ {d \over dx} {1 \over D^2_\pm}  \ .
\eea
Therefore, we use the above result to obtain 
the 4D-massive triple-cut,
\bea
\Theta &\equiv&
{1 \over (2 \pi i)} \Big\{ \Delta_{12}^{(+)} - \Delta_{12}^{(-)} \Big\} 
\nn \\
&=&
- { \spab 1.3.2 \ (1-2z)^3 
\over s_{12} \  z^2(1-z)^2
}
\int dx \ 
{(1-x) x \over 2x - (x_1 + x_2)}
\delta'( D^2 )
\nn \\
&=&
- { \spab 1.3.2 \ (1-2z)^3 
\over s_{12} \  z^2(1-z)^2
}
\int dx \ 
{(1-x) x \over 2x - (x_1 + x_2)}
\nn \\ && \quad \times
\Bigg\{
  { \delta'( x - x_1 ) \over |x_1 - x_2|}
+ { \delta'( x - x_2 ) \over |x_2 - x_1|}
\Bigg\}
\nn \\
&=&
{ \spab 1.3.2 \ (1-2z)^3 
\over s_{12} \  z^2(1-z)^2 \ |x_1 - x_2|
}
  \nn \\ && \quad \times
\int dx \ 
\Big\{
  { \delta( x - x_1 ) }
+ { \delta( x - x_2 ) }
\Big\} \ 
  \nn \\ && \qquad \times
{d \over dx} \ 
{(1-x) x \over 2x - (x_1 + x_2)}
\nn \\
&=&
{ \spab 1.3.2 \ (1-2z)^3 
\over s_{12} \  z^2(1-z)^2 \ |x_1 - x_2|
}
  \nn \\ && \quad \times
\int dx \ 
\Big\{
  { \delta( x - x_1 ) }
+ { \delta( x - x_2 ) }
\Big\} \ 
  \nn \\ && \qquad \times
\frac{-2 x^2+2 \left(x_1 + x_2\right) x-(x_1 + x_2)}
{\left(2 x-(x_1 + x_2)\right)^2}
\nn \\
&=& 
{ (-2) \ 
\over s_{12} \  
} \spab 1.3.2 \ .
\eea
Finally, one uses Eq.(\ref{LinTri:DCUTvs4CUT}), to reconstruct the 
$D$-dimensional triple-cut,
\bea
{\cal N}_{12|3|4} &=& 
\spab 1.3.2 \ 
\chi_{12} \int_0^1 du \ u^{-1 - \epsilon} \ 
\mu^2 \ 
{ (-2) \ 
\over s_{12} \  
} \ ,
\eea
out of which one can read the coefficient $c_3 = \spab 1.3.2$,
multiplying the cut of a $1m$-triangle, 
see Eq.(\ref{finalTripleCutTriangle1m}),
in shifted dimensions, namely $J_3$.
The presence of terms like $\mu^{2 m}$ is sterile for 
the 4D-integration, and it only affects the $u$-integral, 
see Eq.(\ref{eq:udef}). As already said, the integration over $u$
can be performed implicitly, by 
absorbing it in the re-definition of the 
integration measure for a value of dimensions which are shifted from 
the original one, $D \to D + 2m$. That simply translates into the definition 
of the $n$-point $J$-type scalar integral, $J_n \equiv I_n[\mu^2]$, as having
the same denominators as $I_n$ and a single power of $\mu^2$ up in the
numerator \cite{Bern:1995db,Brandhuber:2005jw}.

\subsection{Linear Box}
\label{sec:Linear0mBox}

We consider the linear box integral,
\bea
B^\nu &=& \int d^4 L_2 \ d^{-2 \epsilon}\mu \ 
{\mu^2 L_2^\nu \over
D_1 \ D_2 \ D_3 \ D_4
}
\eea
where
\bea
D_1 &=&  L_2^2 - \mu^2 \ ; \\
D_2 &=& (L_2 + K_{12})^2 - \mu^2 \ ; \\
D_3 &=& (L_2 - k_3)^2 - \mu^2 \ ; \\
D_4 &=& (L_2 + k_4)^2 - \mu^2 \ . 
\eea
In particular let us consider the spinor sandwich 
$\spab 1.B.2$ which 
has the same value of $\spab 1.A.2$
appearing in Eq.(3.43) of \cite{Brandhuber:2005jw},
and whose result, obtained by PV-reduction, reads
\bea
\spab 1.B.2 = \spab 1.3.2 \ \gamma \ ,
\label{LinBox:ToBeMatched}
\eea
with 
\bea
\gamma = {1 \over 2 s_{13}} \big( s_{12} J_4 - 2 J_3 \Big) \ .
\eea
(overall factors understood, see \cite{Brandhuber:2005jw} for details).

The above result, written in terms of $1m$-triangle and $0m$-box
master integrals in shifted dimension, respectively $J_3$ and $J_4$,
can be entirely reconstructed from the triple-cut integration.

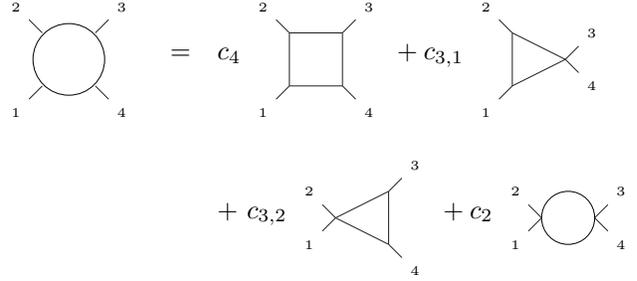
\begin{figure}[h]
\vspace*{0.5cm}
\bea
\hspace*{1.0cm} 
\begin{picture}(0,0)(0,0)
\SetScale{0.5}
\SetWidth{0.5}
\GOval(0,0)(27,27)(0){1}
\Line(-20,-20)(-30,-30)
\Line(20,-20)(30,-30)
\Line(20,20)(30,30)
\Line(-20,20)(-30,30)
%
\Text(-20,-20)[]{\tiny {$1$}}
\Text(-20, 20)[]{\tiny {$2$}}
\Text( 20, 20)[]{\tiny {$3$}}
\Text( 20,-20)[]{\tiny {$4$}}
%
\end{picture}
\hspace*{1cm} &=& 
c_4 \hspace*{1cm} 
\begin{picture}(0,0)(0,0)
\SetScale{0.5}
\SetWidth{0.5}
\Line(-20,-20)(20,-20)
\Line(20,-20)(20,20)
\Line(20,20)(-20,20)
\Line(-20,20)(-20,-20)
\Line(-20,-20)(-30,-30)
\Line(20,-20)(30,-30)
\Line(20,20)(30,30)
\Line(-20,20)(-30,30)
%
\Text(-20,-20)[]{\tiny {$1$}}
\Text(-20, 20)[]{\tiny {$2$}}
\Text( 20, 20)[]{\tiny {$3$}}
\Text( 20,-20)[]{\tiny {$4$}}
%
\end{picture}
\hspace*{1cm} + 
c_{3,1} \hspace*{1cm}
\begin{picture}(0,0)(0,0)
\SetScale{0.5}
\SetWidth{0.5}
\Line(-20,-20)(20,0)
\Line(-20,20)(20,0)
\Line(-20,-20)(-20,20)
\Line(-20,-20)(-30,-30)
\Line(20,0)(30,10)
\Line(20,0)(30,-10)
\Line(-20,20)(-30,30)
%
\Text(-20,-20)[]{\tiny {$1$}}
\Text(-20, 20)[]{\tiny {$2$}}
\Text( 20, 10)[]{\tiny {$3$}}
\Text( 20,-10)[]{\tiny {$4$}}
%
\end{picture}
\nn \\ 
&& \nn \\
&& \nn \\
&& \nn \\
\hspace*{1cm} && 
+ \ c_{3,2} \hspace*{1cm}
\begin{picture}(0,0)(0,0)
\SetScale{0.5}
\SetWidth{0.5}
\Line(-20,0)(20,-20)
\Line(20,-20)(20,20)
\Line(20,20)(-20,0)
\Line(20,-20)(30,-30)
\Line(-20,0)(-30,10)
\Line(-20,0)(-30,-10)
\Line(20,20)(30,30)
%
\Text(-20,-10)[]{\tiny {$1$}}
\Text(-20, 10)[]{\tiny {$2$}}
\Text( 20, 20)[]{\tiny {$3$}}
\Text( 20,-20)[]{\tiny {$4$}}
%
\end{picture}
\hspace*{1cm} + 
c_2 \hspace*{1cm}
\begin{picture}(0,0)(0,0)
\SetScale{0.5}
\SetWidth{0.5}
\GOval(0,0)(20,20)(0){1}
\Line(-20,0)(-30,+10)
\Line(-20,0)(-30,-10)
\Line( 20,0)(30,+10)
\Line( 20,0)(30,-10)
\Text(-20,-10)[]{\tiny {$1$}}
\Text(-20, 10)[]{\tiny {$2$}}
\Text( 20, 10)[]{\tiny {$3$}}
\Text( 20,-10)[]{\tiny {$4$}}
%
\end{picture}
\hspace*{1cm} 
\nn
\eea

\caption{Decomposition of 
a 1-loop 4-point amplitude (or integral) in
  terms of a $0m$-box, two $1m$-triangles and a bubble,
 with rational coefficients $c$'s.}
\label{fig:FourPointDecomposition}
\end{figure}

On general ground, any 4-point integral (therefore any amplitude) admits 
a decomposition in terms of master-integrals, as depicted in
Fig.\ref{fig:FourPointDecomposition}.
The coefficients $c_4, c_{3,1}$, and $c_{3,2}$ can be reconstructed from
triple cuts.
In particular, the integral $\spab 1.B.2$ has two independent
triple-cuts, namely ${\cal N}_{1|2|34}$ 
in Fig.\ref{fig:TripleCut34LinearBox},
and ${\cal N}_{12|3|4}$ in Fig.\ref{fig:TripleCut12LinearBox}, 
which we will discuss separately.

\subsubsection{Triple-cut ${\cal N}_{1|2|34}$}

\begin{figure}[h]
\vspace*{0.5cm}
$$
\begin{picture}(0,0)(0,0)
\SetScale{0.5}
\SetWidth{0.5}
\GOval(0,0)(27,27)(0){1}
\Line(-20,-20)(-30,-30)
\Line(20,-20)(30,-30)
\Line(20,20)(30,30)
\Line(-20,20)(-30,30)
\DashLine(0,32)(0,-32){3}
\DashLine(-30,0)(0,0){3}
\Text(-20,-20)[]{\tiny {$1$}}
\Text(-20, 20)[]{\tiny {$2$}}
\Text( 20, 20)[]{\tiny {$3$}}
\Text( 20,-20)[]{\tiny {$4$}}
\Text(-20, 0)[]{\tiny {$L_1$}}
\Text( 0, 20)[]{\tiny {$L_2$}}
\Text( 20, 0)[]{\tiny {$L_3$}}
\Text( 0,-20)[]{\tiny {$L_4$}}
\end{picture}
\hspace*{1cm} = 
c_4 \hspace*{1cm} 
\begin{picture}(0,0)(0,0)
\SetScale{0.5}
\SetWidth{0.5}
\Line(-20,-20)(20,-20)
\Line(20,-20)(20,20)
\Line(20,20)(-20,20)
\Line(-20,20)(-20,-20)
\Line(-20,-20)(-30,-30)
\Line(20,-20)(30,-30)
\Line(20,20)(30,30)
\Line(-20,20)(-30,30)
\DashLine(0,30)(0,-30){3}
\DashLine(-28,0)(0,0){3}
\Text(-20,-20)[]{\tiny {$1$}}
\Text(-20, 20)[]{\tiny {$2$}}
\Text( 20, 20)[]{\tiny {$3$}}
\Text( 20,-20)[]{\tiny {$4$}}
\Text(-20, 0)[]{\tiny {$L_1$}}
\Text( 0, 20)[]{\tiny {$L_2$}}
\Text( 20, 0)[]{\tiny {$L_3$}}
\Text( 0,-20)[]{\tiny {$L_4$}}
\end{picture}
\hspace*{1cm} + 
c_{3,1}\hspace*{1cm} 
\begin{picture}(0,0)(0,0)
\SetScale{0.5}
\SetWidth{0.5}
\Line(-20,-20)(20,0)
\Line(-20,20)(20,0)
\Line(-20,-20)(-20,20)
\Line(-20,-20)(-30,-30)
\Line(20,0)(30,10)
\Line(20,0)(30,-10)
\Line(-20,20)(-30,30)
\DashLine(0,30)(0,-30){3}
\DashLine(-28,0)(0,0){3}
\Text(-20,-20)[]{\tiny {$1$}}
\Text(-20, 20)[]{\tiny {$2$}}
\Text( 20, 10)[]{\tiny {$3$}}
\Text( 20,-10)[]{\tiny {$4$}}
\Text(-20, 0)[]{\tiny {$L_1$}}
\Text( 0, 20)[]{\tiny {$L_2$}}
\Text( 0,-20)[]{\tiny {$L_4$}}
\end{picture}
$$
\caption{
A triple-cut of a linear box in terms of the master triple-cuts.
}
\label{fig:TripleCut34LinearBox}
\end{figure}
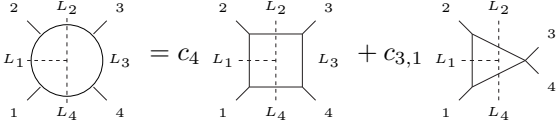

The triple-cut integral corresponding to the
{\it l.h.s.} of Fig.\ref{fig:TripleCut34LinearBox}, is defined as,
\bea
{\cal N}_{1|2|34} &=& 
\int d^D \Phi \  
\delta((L_4 - k_1)^2 - \mu^2) {\mu^2 \ \spab 1.L_4.2 \over (L_4 + k_4) - \mu^2}
\nn \\
&=&
\int d^D \Phi \  
\delta(2 L_4 \cdot k_1 ) {\mu^2 \ \spab 1.L_4.2 \over (2 L_4 \cdot k_4)}
\nn \\
&=&
\chi_{34} \int_0^1 du \ u^{-1 - \epsilon} \ \mu^2 \ \Theta_{1|2|34} 
\label{LinBox:DCUTvs4CUT}
\eea
where
\bea
\Theta_{1|2|34}
&=&
{1 \over (2 \pi i)} \Big\{ \Delta_{34}^{(+)} - \Delta_{34}^{(-)} \Big\} 
\eea
with 
\bea
\Delta_{34}^{(\pm)} &=&
\int d^4\Phi 
{\spab 1.L_4.2 \over (2 L_4 \cdot k_1 )_\pm \ (2 L_4 \cdot k_4)}
\nn \\
&=& 
\int dz {\delta(z-z_0) \over s_{34}(1-2z)}
\int \dea \deb
 \nn \\ &&  \times
\int t^2 dt \ 
\delta \Bigg(t - {(1-2z)s_{34} \over \spab \ell.K_{34}.\ell} \Bigg)
\nn \\ && \qquad \times 
{\spab 1.L_4.2 \over (2 L_4 \cdot k_1 )_\pm \ (2 L_4 \cdot k_4)} 
\ . 
\eea

Since,
\bea
2 L_4 \cdot k_1 &=& {t \over (1-2z)} \spab \ell.Q_1.\ell \ ,
\\
2 L_4 \cdot k_4 &=& {t \over (1-2z)} \spab \ell.Q_2.\ell \ ,
\eea
with
\bea
\s{Q}_1 &\equiv&
(1-2z) \s{k}_1  + z \s{K}_{12} = 
(1-z) \s{k}_1  + z \s{k}_2 \ , 
\\ 
\s{Q}_2 &\equiv&
(1-2z) \s{k}_4  + z \s{K}_{34} = 
(1-z) \s{k}_4  + z \s{k}_3 \ , 
\eea
one can write,
\bea
\Delta_{34}^{(\pm)}
&=& 
\int dz {\delta(z-z_0) \over s_{34}(1-2z)}
\int \dea \deb
\nn \\ && \times 
\int t^2 dt \ 
\delta \Bigg(t - {(1-2z)s_{34} \over \spab \ell.K_{34}.\ell} \Bigg)
\nn \\ && \qquad \times 
{t 
\spab 1.\ell.2 
\over 
{t \over (1-2z)} \spab \ell.Q_1.\ell_\pm \  
{t \over (1-2z)} \spab \ell.Q_2.\ell
}
\nn \\
&=&
\int dz \ \delta(z-z_0) (1-2z)^2 \ {\cal I}_{34}
\ ,
\eea
where
\bea
{\cal I}_{34} &=&
\int \dea \deb
{ 
\spab 1.\ell.2 
\over 
\spab \ell.K_{34}.\ell \ 
\spab \ell.Q_1.\ell_\pm \  
\spab \ell.Q_2.\ell
}
\nn \\
\eea
By means of Schouten identities, and trivial spinor algebra, one can write,
\bea
{\cal I}_{34} &=&
\int \dea \deb
\bigg(
  \nn \\ &&
{
\spa 1.\ell
\spb 3.2
\over
(1-2z)
\spa 4.\ell
\spab \ell.K_{34}.\ell
\spab \ell.Q_1.\ell
\spb 3.4
}
  \nn \\ &&
- 
{z
\spa 1.\ell
\spb 3.2
\over
(1-2z)
\spa 4.\ell
\spab \ell.Q_1.\ell
\spab \ell.Q_2.\ell
\spb 3.4
}
  \nn \\ &&
+
{
\spa 1.\ell
\spb 4.2
\over
(1-2z)
\spa 3.\ell
\spab \ell.K_{34}.\ell
\spab \ell.Q_1.\ell
\spb 3.4
}
  \nn \\ &&
-
{(1-z)
\spa 1.\ell
\spb 4.2
\over
(1-2z)
\spa 3.\ell
\spab \ell.Q_1.\ell
\spab \ell.Q_2.\ell
\spb 3.4
}
\bigg)
\nn \\
&=&
\dex
\int \dea \deb
\bigg(
  \nn \\ &&
-
{
\spa 1.\ell
\spb 3.2
\over
(1-2z)
\spa 4.\ell
\spab \ell.S_1.\ell^2
\spb 3.4
}
  \nn \\ &&
+
{z
\spa 1.\ell
\spb 3.2
\over
(1-2z)
\spa 4.\ell
\spab \ell.R.\ell^2
\spb 3.4
}
  \nn \\ &&
-
{
\spa 1.\ell
\spb 4.2
\over
(1-2z)
\spa 3.\ell
\spab \ell.S_1.\ell^2
\spb 3.4
}
  \nn \\ &&
+
{(1-z)
\spa 1.\ell
\spb 4.2
\over
(1-2z)
\spa 3.\ell
\spab \ell.R.\ell^2
\spb 3.4
}
\bigg) \ ,
\eea
with
\bea
\s{S}_1 &=& - (1-x) \s{K}_{34} + x \s{Q}_1 \ ,
\label{linbox:S1def}
\\
\s{R} &=& x \s{Q}_1  - (1-x) \s{Q}_2 \ .
\label{linbox:Rdef}
\eea

We can separate ${\cal I}_{34}$ into two terms, 
\bea
{\cal I}_{34} &=& {\cal I}_{34}^{(1)} + {\cal I}_{34}^{(2)} 
\eea
where
\bea
{\cal I}_{34}^{(1)} &=&
\dex
\int \dea \deb
\bigg(
  \nn \\ &&
-
{
\spa 1.\ell
\spb 3.2
\over
(1-2z)
\spa 4.\ell
\spab \ell.S_1.\ell^2
\spb 3.4
}
  \nn \\ &&
-
{
\spa 1.\ell
\spb 4.2
\over
(1-2z)
\spa 3.\ell
\spab \ell.S_1.\ell^2
\spb 3.4
}
\bigg) \ ,
\label{I341def}
\\
{\cal I}_{34}^{(2)} &=&
\dex
\int \dea \deb
\bigg(
  \nn \\ &&
+
{z
\spa 1.\ell
\spb 3.2
\over
(1-2z)
\spa 4.\ell
\spab \ell.R.\ell^2
\spb 3.4
}
  \nn \\ &&
+
{(1-z)
\spa 1.\ell
\spb 4.2
\over
(1-2z)
\spa 3.\ell
\spab \ell.R.\ell^2
\spb 3.4
}
\bigg) \ ,
\label{I342def}
\eea
each of which, being characterized by the presence of
either $S_1$ or $R$ , will lead unequivocally to triangle- and box-term
respectively.
Let's, therefore, discuss them separately. \\

\noindent
{\bf{$\bullet \ {\cal I}_{34}^{(1)}$ term}} 

\noindent
To simplify the spinor integration, we use
\bea
{\deb \over \spab \ell.S_1.\ell^2}
&=&
\dedeb {\spb 4.\ell \over \spab \ell.S_1.\ell \spab \ell.S_1.4}
\eea
for the first term of ${\cal I}_{34}^{(1)}$, and
\bea
{\deb \over \spab \ell.S_1.\ell^2}
&=&
\dedeb { \spb 3.\ell \over \spab \ell.S_1.\ell \spab \ell.S_1.3}
\eea
for the second one, yielding
\bea
{\cal I}_{34}^{(1)} 
&=&
\dex {1 \over (S_1^2)}
\bigg(
-
{
\spab 1.S_1.4
\spb 3.2
\over
(1-2z)
\spab 4.S_1.4
\spb 3.4
}
  \nn \\ && \qquad 
-
{
\spab 1.S_1.3
\spb 4.2
\over
(1-2z)
\spab 3.S_1.3
\spb 3.4
}
\bigg)
\eea
Since,
\bea
\spab 1.S_1.4 &=& -(1 - x + x z) \spab 1.3.4 \ , \\ 
\spab 4.S_1.4 &=&  s_{23} (1 + A - A x - x z + A x z) \ , \\  
\spab 1.S_1.3 &=& -(1 - x + x z) \spab 1.4.3 \ , \\  
\spab 3.S_1.3 &=& -s_{23} (-1 - A + x - x z + A x z) \ , 
\eea
with 
\bea
A &=& {s_{24} \over s_{23}} \ ,
\eea
one has,
\bea
{\cal I}_{34}^{(1)} 
&=&
{\spab 1.3.2 \over s_{23} (1 - 2 z) }
\dex \ {f_1(x) \over (S_1^2)} 
\eea
with
\bea
f_1(x) &=& (1 - x + x z)
\nn \\ && \times 
\bigg[ 
{ 1
\over 
 (1 + A - A x - x z + A x z)
}
\nn \\ && \quad 
-
{ 1
\over 
(-1 - A + x - x z + A x z)
}
\bigg] \ ,
\eea
where we used, $\spab 1.4.2 = - \spab 1.3.2$, due to momentum conservation.
Since, see Eq.(\ref{Rroots4Triangle1m}),
\bea
S_1^2 = s_{34} \ z(1-z) \ (x - x_1) (x - x_2) \ ,
\eea
and
\bea
x_1 &=& {1\over z} \ ; \qquad x_2 = {1 \over 1-z} \ , 
\eea
the contribution to the 4D-massive triple-cut reads,
\bea
\delta {\cal I}_{34}^{(1)} 
&=&
{\spab 1.3.2 \over s_{23} \ s_{34} \ z(1-z) (1 - 2 z) }
  \nn \\ && \times 
  \int dx \ f_1(x) \ \delta( (x-x_1) (x-x_2) )
\nn \\
&=&
- { (1+A) \over A \ s_{23} \ s_{34} } \spab 1.3.2
\nn \\
&=&
{\spab 1.3.2 \over \ s_{23} \ s_{34}}
\label{LinBox:deI341final}
\eea
where we used $s_{23} + s_{24} = - s_{34}$, due to momentum conservation. \\

\noindent
{\bf{$\bullet \ {\cal I}_{34}^{(2)}$ term}}

\noindent
The integral ${\cal I}_{34}^{(2)}$ has been defined in (\ref{I342def}).
We use
\bea
{\deb \over \spab \ell.R.\ell^2}
&=&
\dedeb {\spb 4.\ell \over \spab \ell.R.\ell \spab \ell.R.4}
\eea
in the first term of ${\cal I}_{34}^{(2)}$, and
\bea
{\deb \over \spab \ell.R.\ell^2}
&=&
\dedeb { \spb 3.\ell \over \spab \ell.R.\ell \spab \ell.R.3}
\eea
in the second one, yielding
\bea
{\cal I}_{34}^{(2)} 
&=&
\dex {1 \over (R^2)}
\bigg[
{ z \spab 1.R.4 \spb 3.2 
\over
(1-2z) \spab 4.R.4 \spb 3.4}
 \nn \\ && \quad
+
{ (1-z) \spab 1.R.3 \spb 4.2 
\over
(1-2z) \spab 3.R.3 \spb 3.4}
\bigg] \ .
\eea
Since,
\bea
\spab 1. R. 4 &=& -z \spab 1. {3}. 4 \ , \\  
\spab 4. R. 4 &=& -s_{23} (-x - z - A z + 2 x z) \ , \\  
\spab 1. R. 3 &=& -(1 - x - z + 2 x z) \spab 1.{4}. 3 \ , \\  
\spab 3. R. 3 &=& -s_{23} (-1 - A + x + z + A z - 2 x z) \ , \quad 
\eea
one has,
\bea
{\cal I}_{34}^{(2)} 
&=&
{\spab 1.3.2 \over s_{23} \ (1-2z)} 
\dex {f_2(x) \over (R^2)} \ ,
\eea
with
\bea
f_2(x) &=& 
{ z^2 \over (-x - z - A z + 2 x z) }
 \nn \\ && 
+
{ (1 - z) (1 - x - z + 2 x z) \over (-1 - A + x + z + A z - 2 x z)}
\ , 
\eea
where we used, $\spab 1.4.2 = - \spab 1.3.2$.

Since, see Eqs.(\ref{Rroots4Box}),
\bea
R^2 &=& 
s_{23} (1 - 2 z)^2 \ (x-y_1) (x-y_2) \ ,
\eea
where
\bea
y_{1,2} = \frac{1}{2} \left(1 \pm \frac{\sqrt{A u+1}}{\sqrt{1-u}}\right)
\ , 
\qquad
A = { s_{13} \over s_{23} } = { s_{24} \over s_{23} } \ , 
\label{linbox:RRdefs}
\eea
the contribution to the 4D-massive triple-cut reads,
\bea
\delta {\cal I}_{34}^{(2)} 
&=&
{\spab 1.3.2 \over s_{23}^2 \ (1-2z)^3} 
\int dx \ f_2(x) \ \delta((x-y_1) (x-y_2)) \ ,
\nn \\
&=&
- {\spab 1.3.2 \over s_{23} \ s_{24} \ \sqrt{1+Au}} \ .
\label{LinBox:deI342final}
\eea
By combining it with (\ref{LinBox:deI341final}),
one can now write down the result for the 4D-massive triple-cut,
\bea
\Theta_{1|2|34} &=&
\Big(\delta {\cal I}_{34}^{(1)} + \delta {\cal I}_{34}^{(2)}\Big) \ .
\eea

Finally, one uses Eq.(\ref{LinBox:DCUTvs4CUT})
to reconstruct the $D$-dimensional triple-cut,
\bea
{\cal N}_{1|2|34} &=& 
\chi_{12} \int_0^1 du \ u^{-1 - \epsilon} \ \mu^2 \ 
\nn \\ && \times
\bigg\{
  {\spab 1.3.2 \over \ s_{23} \ s_{34}}
- {\spab 1.3.2 \over s_{23} \ s_{24} \ \sqrt{1+Au}}
\bigg\}
\eea
out of which one can read the coefficient
\bea 
c_{3,1} &=& - {\spab 1.3.2 \over 2 \ s_{23}} \ ,
\label{LinBox:c31result}
\eea 
multiplying the cut of a $1m$-triangle in shifted dimensions,
namely $J_3$, see Eq.(\ref{finalTripleCutTriangle1m});
and the coefficient
\bea
c_4 &=& {\spab 1.3.2 \ s_{34} \over 2 \ s_{24}} \ ,
\label{LinBox:c4result}
\eea 
multiplying the cut of a $0m$-box in shifted dimensions,
namely $J_4$,  
see Eq.(\ref{finalTripleCutBox0m}).

\subsubsection{Triple-cut ${\cal N}_{12|3|4}$}

\begin{figure}[h]
\vspace*{0.5cm}
$$
\begin{picture}(0,0)(0,0)
\SetScale{0.5}
\SetWidth{0.5}
\GOval(0,0)(27,27)(0){1}
\Line(-20,-20)(-30,-30)
\Line(20,-20)(30,-30)
\Line(20,20)(30,30)
\Line(-20,20)(-30,30)
\DashLine(0,32)(0,-32){3}
\DashLine(0,0)(30,0){3}
\Text(-20,-20)[]{\tiny {$1$}}
\Text(-20, 20)[]{\tiny {$2$}}
\Text( 20, 20)[]{\tiny {$3$}}
\Text( 20,-20)[]{\tiny {$4$}}
\Text(-20, 0)[]{\tiny {$L_1$}}
\Text( 0, 20)[]{\tiny {$L_2$}}
\Text( 20, 0)[]{\tiny {$L_3$}}
\Text( 0,-20)[]{\tiny {$L_4$}}
\end{picture}
\hspace*{1cm} = 
c_4 \hspace*{1cm} 
\begin{picture}(0,0)(0,0)
\SetScale{0.5}
\SetWidth{0.5}
\Line(-20,-20)(20,-20)
\Line(20,-20)(20,20)
\Line(20,20)(-20,20)
\Line(-20,20)(-20,-20)
\Line(-20,-20)(-30,-30)
\Line(20,-20)(30,-30)
\Line(20,20)(30,30)
\Line(-20,20)(-30,30)
\DashLine(0,30)(0,-30){3}
\DashLine(0,0)(28,0){3}
\Text(-20,-20)[]{\tiny {$1$}}
\Text(-20, 20)[]{\tiny {$2$}}
\Text( 20, 20)[]{\tiny {$3$}}
\Text( 20,-20)[]{\tiny {$4$}}
\Text(-20, 0)[]{\tiny {$L_1$}}
\Text( 0, 20)[]{\tiny {$L_2$}}
\Text( 20, 0)[]{\tiny {$L_3$}}
\Text( 0,-20)[]{\tiny {$L_4$}}
\end{picture}
\hspace*{1cm} + 
c_{3,2}\hspace*{1cm} 
\begin{picture}(0,0)(0,0)
\SetScale{0.5}
\SetWidth{0.5}
\Line(-20,0)(20,-20)
\Line(20,-20)(20,20)
\Line(20,20)(-20,0)
\Line(20,-20)(30,-30)
\Line(-20,0)(-30,10)
\Line(-20,0)(-30,-10)
\Line(20,20)(30,30)
\DashLine(0,30)(0,-30){3}
\DashLine(0,0)(28,0){3}
\Text(-20,-10)[]{\tiny {$1$}}
\Text(-20, 10)[]{\tiny {$2$}}
\Text( 20, 20)[]{\tiny {$3$}}
\Text( 20,-20)[]{\tiny {$4$}}
%
\Text( 0, 20)[]{\tiny {$L_2$}}
\Text( 20, 0)[]{\tiny {$L_3$}}
\Text( 0,-20)[]{\tiny {$L_4$}}
\end{picture}
$$
\caption{
A triple-cut of a linear box in terms of the master triple-cuts.
}
\label{fig:TripleCut12LinearBox}
\end{figure}
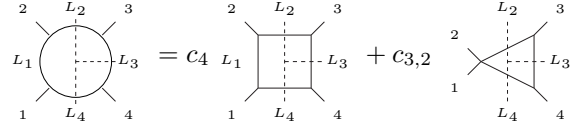

The second triple-cut needed for the reconstruction of the
linear box-integral $\spab1.B.2$ is depicted in
to the {\it l.h.s.} of Fig.\ref{fig:TripleCut12LinearBox}, 
and is defined as,
\bea
{\cal N}_{12|3|4} &=& 
\int d^D \Phi \  
\delta((L_4 + k_4) - \mu^2) {\mu^2 \ \spab 1.L_4.2 \over (L_4 - k_1)^2 - \mu^2}
\nn \\
&=&
\chi_{12} \int_0^1 du \ u^{-1 - \epsilon} \ \mu^2 \ \Theta_{12|3|4} 
\eea
where
\bea
\Theta_{12|3|4}
&=&
{1 \over (2 \pi i)} \Big\{ \Delta_{12}^{(+)} - \Delta_{12}^{(-)} \Big\} 
\eea
with 
\bea
\Delta_{12}^{(\pm)} &=&
\int dz \ \delta(z-z_0) (1-2z)^2 \ {\cal I}_{12}
\ ,
\eea
where
\bea
{\cal I}_{12} &=&
\int \dea \deb
{ 
\spab 1.\ell.2 
\over 
\spab \ell.K_{34}.\ell \ 
\spab \ell.Q_1.\ell \  
\spab \ell.Q_2.\ell_\pm
}
\nn \\
\eea
By means of Schouten identities, different from the ones used 
for ${\cal N}_{1|2|34}$, one can write,
\bea
{\cal I}_{12} &=&
\int \dea \deb
\bigg(
- 
{ \spa 1.\ell \over 
(1-2z) 
\spa 2.\ell 
\spab \ell.K_{34}.\ell 
\spab \ell.Q_2.\ell 
}
\nn \\ && \quad
-
{ (1-z) \spa 1.\ell \over 
(1-2z) 
\spa 2.\ell 
\spab \ell.Q_1.\ell 
\spab \ell.Q_2.\ell 
}
\bigg)
\nn \\ 
&=&
\dex
\int \dea \deb
\bigg(
- 
{ \spa 1.\ell \over 
(1-2z) 
\spa 2.\ell 
\spab \ell.S_2.\ell^2 
}
\nn \\ && \quad
+
{ (1-z) \spa 1.\ell \over 
(1-2z) 
\spa 2.\ell 
\spab \ell.R.\ell^2
}
\bigg) \ ,
\eea
where $\s{R}$ has been defined in Eq.(\ref{linbox:Rdef}), and 
\bea
\s{S}_2 &=&  (1-x) \s{K}_{34} + x \s{Q}_2 \ ,
\label{linbox:S2def}
\eea

We can separate ${\cal I}_{12}$ into two terms, 
\bea
{\cal I}_{12} &=& {\cal I}_{12}^{(1)} + {\cal I}_{12}^{(2)} 
\eea
where
\bea
{\cal I}_{12}^{(1)}
&=&
- \dex
\int \dea \deb
{ \spa 1.\ell \over 
(1-2z) 
\spa 2.\ell 
\spab \ell.S_2.\ell^2 
} 
\ ,
\label{linbox:I121def}
\\ 
{\cal I}_{12}^{(2)} 
&=&
\dex
\int \dea \deb
{ (1-z) \spa 1.\ell \over 
(1-2z) 
\spa 2.\ell 
\spab \ell.R.\ell^2
}
\ ,
\label{linbox:I122def}
\eea
each of which, being characterized by the presence of
either $S_2$ or $R$ , will lead unequivocally to triangle- and box-term
respectively.
Let's, therefore, discuss them separately. \\

\noindent
{\bf{$\bullet \ {\cal I}_{12}^{(1)}$ term}}

\noindent
To simplify the spinor integration we use
\bea
{\deb \over \spab \ell.S_2.\ell^2}
&=&
\dedeb {\spb 2.\ell \over \spab \ell.S_2.\ell \spab \ell.S_1.2}
\eea
to write
\bea
{\cal I}_{12}^{(1)}
&=&
- \dex { \spab 1.S_2.2 \over (1-2z) \spab 2.S_2.2 \ (S_2^2) } \ .
\eea
Since,
\bea
\spab 1. S_2. 2 &=& x (-1 + 2 z) \spab 1.{3}.2 \ , 
\\ 
\spab 2. S_2. 2 &=& -s_{23} (-1 - A + x - x z + A x z) \ ,
\eea
one gets,
\bea
{\cal I}_{12}^{(1)}
&=& 
- {\spab 1.3.2 \over s_{23}}
\dex {f_3(x) \over (S_2^2)} \ ,
\eea
where
\bea
f_3(x) &=& {1 \over (-1 - A + x - x z + A x z)} \ . 
\eea
Since,
\bea
S_2^2 = s_{34} \ z(1-z) \ (x - x_1) (x - x_2) \ ,
\eea
and
\bea
x_1 &=& {1\over z} \ ; \qquad x_2 = {1 \over 1-z} \ , 
\eea
the contribution to the 4D-massive triple-cut reads,
\bea
\delta {\cal I}_{12}^{(1)} 
&=&
- {\spab 1.3.2 \over s_{23} \ s_{34} \ z(1-z)}
 \nn \\ && \quad \times 
\int dx \ f_3(x) \ \delta((x - x_1) (x - x_2)) \ ,
\nn \\
&=&
{(-1 + A) \spab 1.3.2 \over A \ s_{23} \ s_{34}
}
\nn \\ 
&=&
{
(s_{34} - s_{23}) \spab 1.3.2
\over
s_{23} \ s_{24} \ s_{34} \ .
}
\label{LinBox:deI121final}
\eea

\noindent
{\bf{$\bullet \ {\cal I}_{12}^{(2)}$ term} }

\noindent
The integral ${\cal I}_{12}^{(2)}$ has been defined in (\ref{linbox:I122def}).
In this case, we use as well,
\bea
{\deb \over \spab \ell.R.\ell^2}
&=&
\dedeb {\spb 2.\ell \over \spab \ell.R.\ell \spab \ell.R.2}
\eea
to have,
\bea
{\cal I}_{12}^{(2)}
&=&
\dex { (1-z) \spab 1.R.2 \over (1-2z) \spab 2.R.2 \ (R^2) } \ .
\eea
Since,
\bea
\spab 1.R.2 &=& (1 - x) (1 - 2 z) \spab 1.{3}.2 \ , 
\\ 
\spab 2.R.2 &=& s_{23} (-A - x - z + A z + 2 x z) \ ,
\eea
one gets,
\bea
{\cal I}_{12}^{(2)}
&=& 
{ (1 - z) \spab 1.{3}.2 \over s_{23}}
\dex \ {f_4(x) \over (R^2)} \ ,
\eea
where
\bea
f_4(x) &=& 
{(1 - x) \over (-A - x - z + A z + 2 x z)} \ .
\eea
Therefore, given the expression of $R^2$ in
(\ref{linbox:RRdefs}), the contribution to the 4D-massive triple-cut reads,
\bea
\delta {\cal I}_{12}^{(2)} 
&=&
{ (1 - z) \spab 1.{3}.2 \over s_{23}^2 (1-2z)^2}
\int dx \ f_4(x) \delta((x-y_1) (x-y_2)) 
\nn \\ 
&=&
- { \spab 1.3.2 \over A \ s_{23}^2 \ \sqrt{1+Au}}
\nn \\ 
&=&
- { \spab 1.3.2 \over \ s_{23} s_{24} \ \sqrt{1+Au}}
\ .
\label{LinBox:deI122final}
\eea
By combining it with (\ref{LinBox:deI121final}),
one can now write down the result for the 4D-massive triple-cut,
\bea
\Theta_{12|3|4} &=&
\Big(\delta {\cal I}_{12}^{(1)} + \delta {\cal I}_{12}^{(2)}\Big) \ .
\eea

Finally, one uses Eq.(\ref{LinBox:DCUTvs4CUT})
to reconstruct the $D$-dimensional triple-cut,
\bea
{\cal N}_{1|2|34} &=& 
\chi_{12} \int_0^1 du \ u^{-1 - \epsilon} \ \mu^2 \ 
\nn \\ && \times
\bigg\{
  {\spab 1.3.2 \ (s_{34} - s_{23}) \over s_{23} \ s_{24} \ s_{34} }
- { \spab 1.3.2 \over \ s_{23} s_{24} \ \sqrt{1+Au}}
\bigg\}
\nn \\ 
\eea
out of which one can read the coefficient
\bea 
c_{3,2} &=& - {\spab 1.3.2 \ (s_{34} - s_{23}) \over 2 \ s_{23} \ s_{24}} \ ,
\label{LinBox:c32result}
\eea 
multiplying the cut of a $1m$-triangle in shifted dimensions,
namely $J_3$, see Eq.(\ref{finalTripleCutTriangle1m});
and the coefficient
\bea
c_4 &=& {\spab 1.3.2 \ s_{34} \over 2 \ s_{24}} \ ,
\eea 
multiplying the cut of a $0m$-box in shifted dimensions,
namely $J_4$,  
see Eq.(\ref{finalTripleCutBox0m}).
We notice, that, as it should be, $c_4$ extracted from the triple-cut
${\cal N}_{12|3|4}$ is the same as obtained in
(\ref{LinBox:c4result}) from ${\cal N}_{1|2|34}$.

The matching with the result of \cite{Brandhuber:2005jw}
re-written here in (\ref{LinBox:ToBeMatched}) can
be confirmed. The coefficient of $J_4$ 
in (\ref{LinBox:ToBeMatched})
is exactly our $c_4$, since
$s_{34}=s_{12}$ and $s_{24}=s_{13}$. Whereas the coefficient
of $J_3$ in (\ref{LinBox:ToBeMatched}) amounts to the sum 
$(c_{3,1} + c_{3,2})$, because, accidentally, the two 1m-triangles 
in Fig.\ref{fig:FourPointDecomposition} can be expressed by the same
function, $J_3$.

\section{Triple-Cut in Four Dimensions}

Given the decomposition of a triple-cut in terms 
of two double-cuts, see Fig.\ref{Fig:TripleCut}, in order to compute
4D-massless triple-cut, one has to use simply the 
two-particle massless phase-space, 
$d^4 \phi$ \cite{Britto:2004nj,Britto:2005ha},
\bea
\int d^4 \phi 
&=&
\int {\dea \deb \over \spab \ell.K.\ell}
\int t \ dt \ 
\delta \Bigg(t - {(K^2 \over \spab \ell.K.\ell}\Bigg) \ ,
\label{phi4massless}
\eea
and perform the spinor integration along the line of 
\cite{Britto:2005ha,Britto:2006sj}.

Triple-cut in four dimension allow the extraction of the 
coefficients of triangle- and box-functions from finite cuts
of one-loop amplitudes, which enable the complete reconstruction of
amplitudes, for example, in Supersymmetry and Gravity.

\section{Conclusions}

We have presented a new method for computing {\it triple cuts} 
of dimensional regularised one-loop amplitudes.
It enables the direct extraction of
triangle- and higher-point-function coefficients from
any one-loop amplitude in arbitrary dimensions.

The triple-cut has been defined as a difference of two double-cuts, 
with the same particle contents and
a same propagator carrying opposite $i0$-prescription
in each of the two cuts.

The three-particle $D$-dimensional phase-space measure is 
written as a standard convolution of a four-dimensional massive 
three-particle phase-space, and an 
integration over the corresponding mass parameter, which 
plays the role of the $(-2\epsilon)$-dimensional scale.

The four-dimensional integration, in each double-cut, 
is carried on as, together with Anastasiou, Britto, Feng, and Kunszt, 
we have recently proposed, 
by combining the method of spinor integration via the holomorphic anomaly
of massive phase-space integrals, and an integration over the 
Feynman parameter. 
After Feynman parametrisation, by combining back the two double-cuts
into the triple-cut,
the parametric integration is reduced to the extraction of residues
to the branch-points in correspondence of the zeroes of a standard function
of the Feynman parameter, hereby called SQF'n,
characterizing each master-integral.
The final integration over the dimensional scale parameter
is mapped directly to triple-cut of 
master integrals with shifted dimensions. 

Along the line of the Feynman Tree Theorem for reconstructing any amplitude 
from its multiple generalised discontinuities,
one can now compute $n$-point ($n \ge 4$) coefficients from quadruple cuts,
three-point coefficients from triple-cuts, 
and two-point coefficients from double-cuts, by avoiding 
the conventional tensor reduction. 
Thus, given the decomposition of any amplitude in terms of MI, 
the coefficient of any $n$-point MI
can be recovered from the all-channels $n$-particle cut.
Each $n$-particle cut may detect as well higher-point MI, 
which will appear with different analytic structure,
for it comes from specific zeroes of a standard quadratic function 
of the Feynman parameter.

The triple-cut method hereby outlined 
can be applied to scattering amplitudes in gauge
theories, and in Gravity as well. In particular, in the latter case, 
it could be employed for the analytical 
investigation of the so called ``no-triangle hypothesis'' of ${\cal N}=8$
Supergravity amplitudes, conjectured by Bern, Bjerrum-Bohr and Dunbar.

On the more speculative side,
we think that the characterization of master integrals in terms of the zeroes
of the corresponding SQF'n
might lead to a deeper understanding of the decomposition of one-loop 
amplitudes in terms of basic scalar integrals, and, possibly, of their 
recursive behaviour.

\section*{Acknowledgments}

This work is supported by Marie-Curie EIF under the contract
MEIF-CT-2006-024178.
We whish to thank 
Mario Argeri, Nigel Glover,
Gudrun Heinrich, and Silvia Pascoli for
stimulating discussions
and in particular Ettore Remiddi for clarifying discussion on 
the analytic properties of multi-particle cuts and the link between
the Feynman Tree Theorem and the Veltman Largest Time Equation. 
We acknowledge the 
hospitality of IPPP at Durham University, and of Bologna Universty, 
where part of this work has been displayed.

\end{document}